\documentclass[USenglish,twocolumn]{article}

\ifx\directlua\undefined\ifx\XeTeXcharclass\undefined
  \usepackage[utf8]{inputenc}                           %pdftex engine
  \else\RequirePackage[no-math]{fontspec}[2017/03/31]\fi %xetex engine
  \else\RequirePackage[no-math]{fontspec}[2017/03/31]\fi %luatex engine
\usepackage[sort&compress,square,numbers]{natbib}
\usepackage[big,online]{dgruyter}

\theoremstyle{dgthm}

\theoremstyle{dgdef}

\begin{document}

\articletype{Letter}

\author[1]{Ilia Sokolovskii}
\author*[2]{Gerrit Groenhof}

\affil[1]{Nanoscience Center and Department of Chemistry, University of
  Jyv\"{a}skyl\"{a}, P.O. Box 35, 40014 Jyv\"{a}skyl\"{a},
  Finland, ilia.i.sokolovskii@jyu.fi; 0000-0003-3367-0660}
\affil[2]{Nanoscience Center and Department of Chemistry, University of
  Jyv\"{a}skyl\"{a}, P.O. Box 35, 40014 Jyv\"{a}skyl\"{a},
  Finland, gerrit.x.groenhof@jyu.fi; 0000-0001-8148-5334}

\title{Photochemical Initiation of Polariton Propagation}
\runningtitle{Photochemical Initiation of Polariton Propagation}
\abstract{Placing a material inside an optical cavity can enhance transport of excitation energy by hybridizing excitons with confined light modes into polaritons, which have a dispersion that provides these light-matter quasi-particles with low effective masses and very high group velocities. While in experiments polariton propagation is typically initiated with laser pulses, tuned to be resonant either with the polaritonic branches that are delocalized over many molecules, or with an uncoupled higher-energy electronic excited state that is localized on a single molecule, practical implementations of polariton-mediated exciton transport into devices would require operation under low-intensity incoherent light conditions. Here, we propose to initiate polaritonic exciton transport with a photo-acid, which upon absorption of a photon in a spectral range not strongly reflected by the cavity mirrors, undergoes ultra-fast excited-state proton transfer into a red-shifted excited-state photo-product that can couple collectively with a large number of suitable dye molecules to the modes of the cavity. By means of atomistic molecular dynamics simulations we demonstrate that cascading energy from a photo-excited donor into the strongly coupled acceptor-cavity states can indeed induce long-range polariton-mediated exciton transport. }
\journalname{Nanophotonics}
%\dedication{our professor, Dr. Firstname Lastname, whose critique of the underlying study identified potential bias in the analysis and strengthened our argument.}
\journalyear{2023}
\journalvolume{aop}

\maketitle

\section{Introduction} 

Organic opto-electronic materials offer many advantages over their silicon counterparts, such as lower production cost, smaller weight, higher flexibility and easier disposability, but are hampered by low exciton mobility~\citep{Mikhnenko2015}. Enhancing that mobility therefore has become a major optimization target and several solutions have been proposed, which include  increasing the lifetime via triplet formation~\citep{Pope1977,Akselrod2014}, ordering molecules to increase exciton delocalization~\citep{Sneyd2021,Kong2022, Sneyd2022,Hildner2023}, or coupling the excitons to the confined light modes of an optical resonator~\citep{Lerario2017,Rozenman2018,Forrest2020,Pandya2021,Ostrovskaya2021,Berghuis2022,Pandya2022,Xu2022,Balasubrahmaniyam2023}. Because the latter solution does not require chemical modifications of the molecules, which may compromise other properties, utilizing strong light-matter coupling could be a promising route towards improving the performance of organic opto-electronic devices. 

% introduction to polaritons

\begin{figure*}\centering
\includegraphics[width=\textwidth]{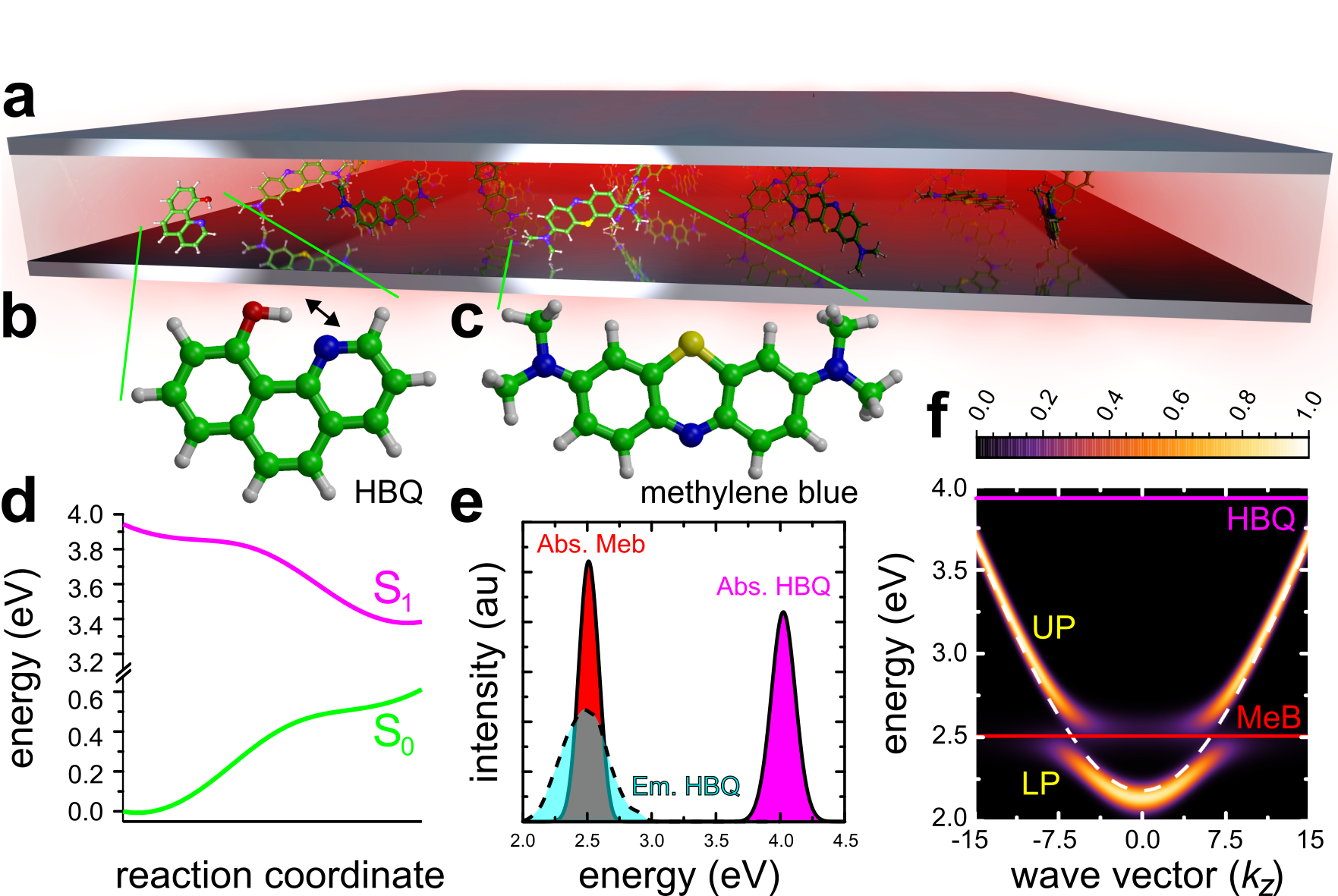}
  \caption{Illustration of a Fabry-P\'{e}rot microcavity (panel {\bf{a}}, not to scale) containing a 10-hydroxybenzo[h]quinoline donor molecule (HBQ, panel {\bf{b}}) and 1023~Methylene Blue acceptor molecules (MeB, panel {\bf{c}}). The first singlet excited states (S$_1$) of the MeB molecules are coupled to the 239 modes of the cavity. Upon absorbing a photon at a frequency where the mirrors have become more transparent ($\sim$ 4.0 eV at the TDA-CAMB3LYP/3-21G level of theory, Figure~S3 in SM), HBQ undergoes ultra-fast intra-molecular proton transfer on the S$_1$ excited-state potential energy surface (panel {\bf{d}}) into an excited-state photo-product that is resonant with both the absorption maximum of MeB and the cavity. Panel {\bf{e}} shows the QM/MM absorption (magenta) and emission (cyan) spectra of HBQ and the absorption spectrum of MeB (red). The normalised angle-resolved absorption spectrum of the molecule-cavity system (panel {\bf{f}}) shows the Rabi splitting of 282 meV between the lower polariton (LP) and upper polariton (UP) branches. The cavity dispersion is plotted as a white dashed line, while the excitation maxima of the MeB molecules ($\sim$ 2.5 eV at the TD-B97/3-21G level of theory) and HBQ are plotted as straight red and magenta lines, respectively.}
\label{fig:overview_system}
\end{figure*}

Because the confinement of light into smaller volumes by an optical resonator increases the interaction with molecular transitions~\citep{Vahala2003}, the enhanced exciton mobility in the strong coupling regime has been attributed to hybridization of excitons and confined light modes into polaritons~\citep{Agranovich2007,Litinskaya2008,Feist2015,Schachenmayer2015,Engelhardt2023,Aroeira2023}, which can form when the interaction strength exceeds the decay rates of both excitons and cavity modes~\citep{Torma2015,Rider2022}. The hybrid states with contributions from cavity modes are bright and can hence be accessed optically~\citep{Agranovich2003,Litinskaya2004}. Because the cavity mode energy depends on the in-plane momentum, or wave-vector, $k_z$, these states have dispersion and form the upper and lower polaritonic branches, as shown in Figure~\ref{fig:overview_system}{\bf{f}}. These branches are separated by the Rabi splitting, which is defined as the energy gap at the wave-vector for which the energy of the exciton and cavity dispersion are resonant. Most of the hybrid states, however, have negligible contribution from the cavity modes, and are hence dark~\citep{Martinez2019}. These dark states therefore also lack dispersion and form a quasi-degenerate manifold instead that is situated in between the two bright polaritonic branches.

% polariton propagation

Owing to their dispersion, the bright polaritonic states support ballistic motion of population at their group velocity (\textit{i.e.}, $v_\text{g}=\partial\omega(k_z)/\partial k_z$, with $\hbar\omega(k_z)$ the energy of a polariton with in-plane momentum $k_z$, Figure~\ref{fig:overview_system}{\bf{f}})~\citep{Agranovich2007, Michetti2008b,Litinskaya2008,Ribeiro2022,Xu2022,Engelhardt2023,Aroeira2023}. However, while in inorganic micro-cavities, such ballistic propagation was indeed observed~\citep{Freixanet2000,Myers2018}, transport in organic micro-cavities is a diffusion process because of rapid dephasing in disordered organic materials~\citep{Litinskaya2008}. Results from molecular dynamics (MD) simulations suggest that such dephasing is due to reversible exchange of population between the stationary dark states and propagating polaritonic states~\citep{Berghuis2022,Sokolovskii2022,Tichauer2023}. Although polariton-mediated exciton transport is not ballistic in organic systems, polaritonic diffusion can still dramatically outperform the intrinsic exciton diffusivity of the material~\citep{Rozenman2018,Balasubrahmaniyam2023}. However, despite several experimental realizations~\citep{Lerario2017,Rozenman2018,Berghuis2022,Pandya2022,Balasubrahmaniyam2023}, and an emerging theoretical understanding of polariton propagation in organic microcavities~\citep{Agranovich2007,Litinskaya2008,Michetti2008b,Feist2015,Schachenmayer2015,Wellnitz2022,Ribeiro2022,Engelhardt2023,Aroeira2023}, strong light-matter coupling has so far not been leveraged systematically for practical applications.

%figure wavepackets and populations

% goal: improve polariton propagation initiation with ESIPT, and demonstrate by MD

\begin{figure}
\includegraphics[width=\columnwidth]{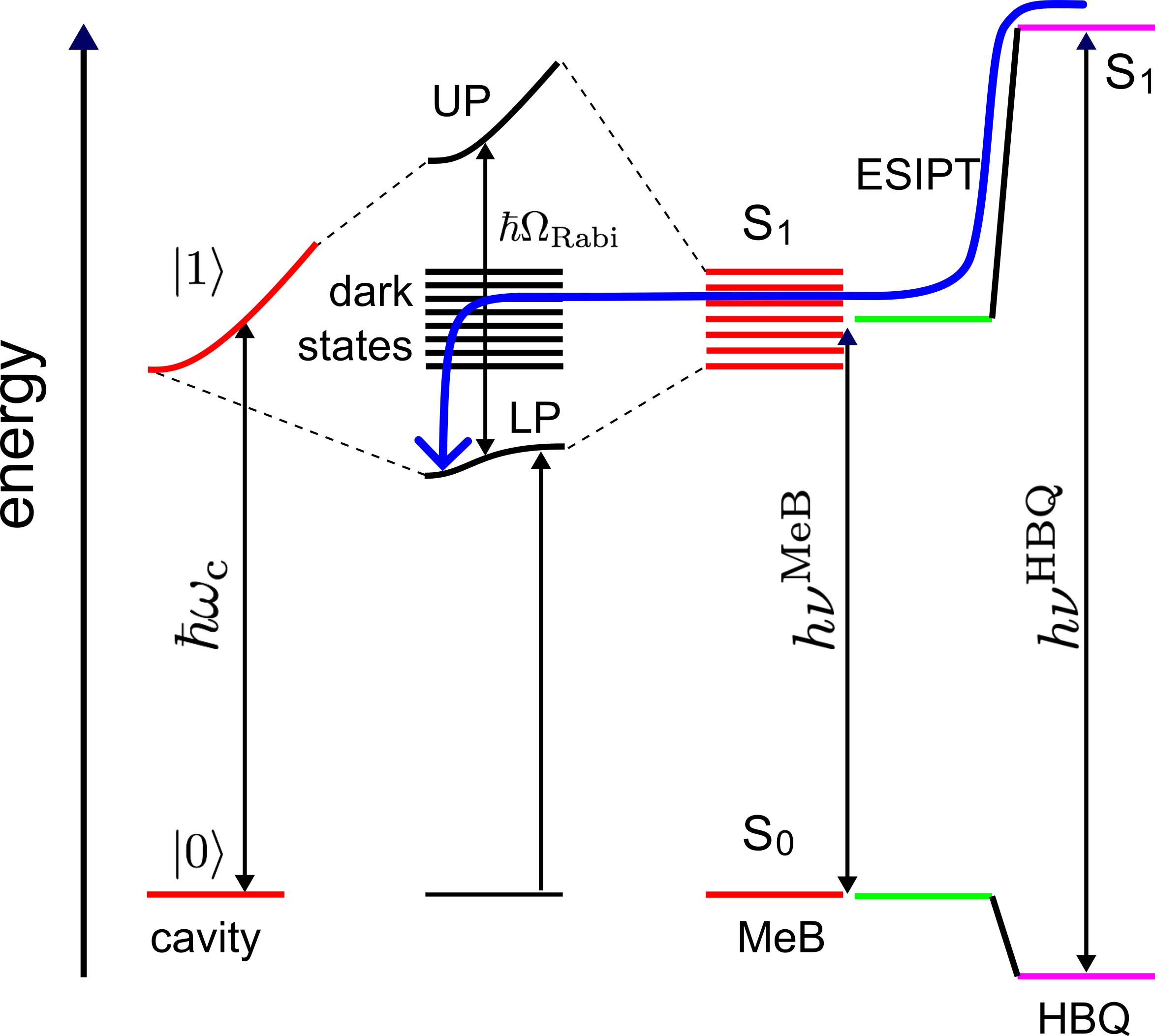}
  \caption{Simplified Jablonski diagram of states involved in photo-chemically induced polariton propagation. The cavity modes and polaritons are schematically shown as continuous dispersions. After photo-excitation of HBQ at $h\nu^\text{HBQ}$, excited-state intra-molecular proton transfer (ESIPT) brings the excited state population that was initially localized on HBQ, into the dark state manifold. Reversible population exchanges between the stationary dark states and the propagating lower polaritonic (LP) bright states, cause the population to move away from the HBQ molecule and diffuse into the cavity~\citep{Sokolovskii2022}. The path along which the population arrives in the LP states is illustrated by a blue arrow.}
\label{fig:Jablonski}
\end{figure}

One of the obstacles on the path to polaritonic devices for enhanced exciton transfer is that polariton propagation requires laser excitation of either wavepackets of polaritonic states~\citep{Pandya2021,Pandya2022}, or higher-energy electronic states of the molecules~\citep{Lerario2017,Rozenman2018, Forrest2020,Balasubrahmaniyam2023}. Yet, for practical applications, such as light-harvesting, it will be essential that transport can also be initiated with low-intensity incoherent light sources. To address this specific challenge for a Fabry-P\'{e}rot optical resonator, we propose to initiate polariton propagation in a strongly coupled molecule-cavity system with a suitable donor that, upon excitation at wavelengths for which the cavity mirrors are more transparent~\citep{Hutchison2012}, undergoes a rapid photo-chemical reaction into an excited-state photo-product with an emission maximum that is resonant with both the cavity and acceptor dye molecules. As illustrated in Figures~\ref{fig:overview_system} and \ref{fig:Jablonski}, such system could potentially be realized if we combine 10-hydroxybenzo[h]quinoline (HBQ) that upon excitation at 375~nm or 360~nm undergoes ultra-fast excited-state intra-molecular proton transfer (ESIPT) on a femtosecond timescale into a photo-product with a broad emission centered at 620~nm~\citep{Kim2009,Lee2013}, with Methylene Blue (MeB) in an optical micro-cavity made of silver mirrors and resonant with the broad absorption peaks of MeB at 668~nm or 609~nm (Figure~S3, in the Supplemantary Material). Here, we demonstrate the feasibility of this concept, which is somewhat similar to radiative pumping of the LP of a strongly-coupled dye with the emission of a weakly coupled second dye~\citep{Akselrod2013}, or to the photo-transformation of an uncoupled reactant into a coupled photo-product~\citep{Hutchison2012,Cargioli2023}, by means of multi-scale molecular dynamics simulations~\citep{Luk2017,Tichauer2021}.

\section{Materials and Methods}

In the simulations, the details of which can be found in the Supplementary Material (SM), a single HBQ molecule solvated in cyclohexane, was combined with 1023~hydrated MeB molecules. While the number of molecules in real cavities is estimated to be much higher (\textit{i.e.}, $10^5-10^8$~\citep{delPino2015,Eizner2019,Martinez2019}), we could show in previous work that for modeling polariton propagation, including 1024~molecules in the simulation provides a reasonable compromise~\citep{Sokolovskii2022}. The electronic ground (S$_0$) and excited (S$_1$) states of HBQ were modeled with Density Functional Theory (DFT)~\citep{Hohenberg1964} and time-dependent density functional theory (TDDFT)~\citep{Runge1984} within the Tamm-Dancoff approximation (TDA)~\citep{Hirata1999}, respectively, using the CAM-B3LYP functional~\citep{Becke1993,Yanai2004} in combination with the 3-21G basis set~\citep{Ditchfield1971}. The cyclohexane solvent molecules were modelled with the GROMOS~2016H66 force field~\citep{Horta2016}. At this level of theory the vertical excitation energy of HBQ is $h\nu^\text{HBQ}=$ 4.06~eV (305~nm), while the energy gap to the ground state is 2.58~eV (480~nm) in the S$_1$ minimum. Despite the overestimation of the S$_1$-S$_0$ energy gap, our model provides potential energy surfaces (Figure~\ref{fig:overview_system}{\bf{d}}) that are in qualitative agreement with the more accurate description at the TPSSh/cc-pVDZ level of theory for this system (Figure~S4 in SM)~\citep{Staroverov2003,Picconi2021}. The S$_0$ and S$_1$ electronic states of MeB were modelled with DFT and TDDFT based on the Casida equations~\citep{Casida1998}, respectively, using the B97 functional~\citep{Becke97} and the 3-21G basis set. The water molecules were described with the TIP3P model~\citep{Jorgensen1983}. Although at this level of theory the vertical excitation energy of MeB is $h\nu^\text{MeB}=$ 2.5~eV, and thus significantly overestimated with respect to experiment, there is a fortuitous overlap with the emission of HBQ that we exploit in this work (Figure~\ref{fig:overview_system}{\bf{e}}). Thus, while MeB may not be the optimal choice for a practical realization, this dye should be suitable for demonstrating the feasibility of inducing polariton-mediated exciton transfer with a photo-chemical reaction in our simulations.

We modelled the optical resonator as a one-dimensional Fabry-P\'{e}rot cavity~\citep{Michetti2005} with 239 discrete modes (Figure~\ref{fig:overview_system}{\bf{f}}) and a cavity vacuum field strength of 0.21~MVcm$^{-1}$ (0.00004~au). The fundamental mode at $k_z = 0$ was red-detuned by 323~meV with respect to the excitation maximum of MeB (2.5~eV at the TD-B97/3-21G level of theory). To maximize light-matter coupling, the transition dipole moments of the molecules were aligned with the vacuum field of the cavity. We computed Ehrenfest MD trajectories, in which the nuclei evolve classically on the mean-field potential energy surface of the total light-matter wave function (SM). To account for the finite lifetime of the cavity modes (here $\tau_\text{cav}$ =~15~fs, in line with experiments on metallic micro-cavities~\citep{Schwartz2013}), the wave function was propagated along the classical trajectory under the influence of an effective non-Hermitian Hamiltonian (SM), in which the loss-terms were added to the cavity mode energies (\textit{i.e.}, $ \hbar\omega(k_z)-1/2i\gamma_\text{cav}$, with $\hbar\omega(k_z)$ the dispersion of the empty cavity, shown in dashed white lines in Figure~\ref{fig:overview_system}{\bf{f}}, and $\gamma_\text{cav}=1/\tau_\text{cav}$ the decay rate of the cavity)~\citep{Ulusoy2020,Antoniou2020,Felicetti2020,Hu2022}. 

\section{Results and Discussion}

% proton transfer, entry into the DS manifold

\begin{figure*}
\centering
\includegraphics[width=\textwidth]{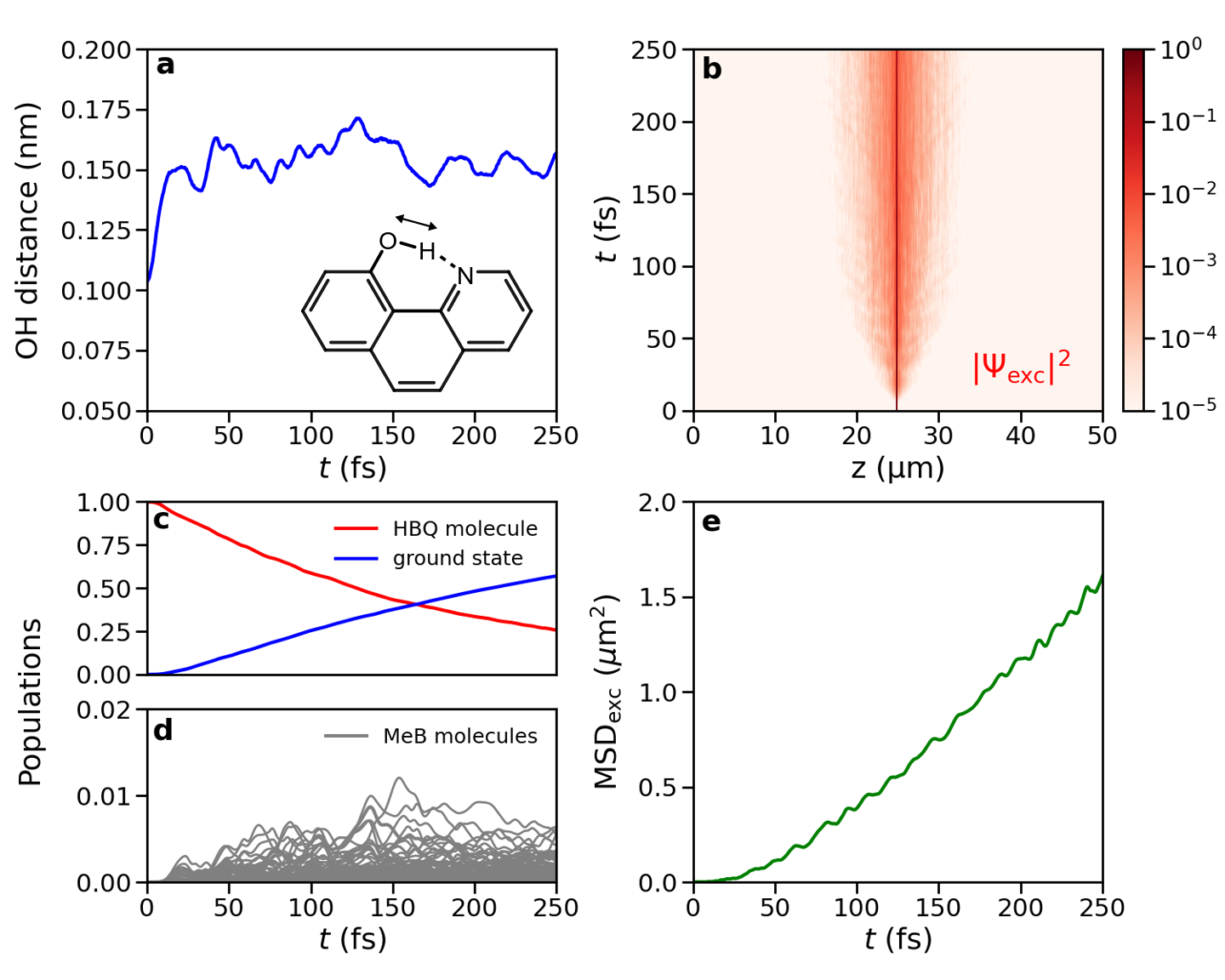}
  \caption{Panel {\bf{a}} shows the distance between the oxygen and proton, which we use as the reaction coordinate for the excited-state proton transfer reaction (inset and see also Figure~\ref{fig:overview_system}{\bf{c}}) as a function of time after instantaneous excitation into the highest energy eigenstate of the molecule-cavity system, which is dominated by the S$_1$ state of HBQ (\textit{i.e.}, $|\beta_{\text{HBQ}}|^2 \approx$ 0.99, panel {\bf{c}}). Panel {\bf{b}} depicts the probability density of the excitonic part of the wave function $|\Psi_{\text{exc}}|^2$ as a function of the $z$-coordinate (horizontal axis) and time (vertical axis). Panels {\bf{c}} and {\bf{d}} show the contribution of the HBQ molecule (red) and the Methylene Blue molecules (grey) to the total wave function, as well as the population of the ground state (blue). Panel {\bf{e}} shows the Mean Squared Displacement (\textit{i.e.}, $\text{MSD}_{\text{exc}} = \langle z(t)-z(0)\rangle^2$) of the excitonic wavepacket.}
\label{fig:transport}
\end{figure*}

In Figure~\ref{fig:transport} we plot the progress of the ESIPT reaction, defined as the distance between the hydroxyl oxygen and the proton ({\bf{a}}), the excitonic wavepacket $|\Psi_{\text{exc}}(z,t)|^2$ ({\bf{b}}), the contributions of the molecular excitations to the total wave function, $|\Psi(z,t)|^2$ ({\bf{c}},\textbf{d}), and the Mean Squared Displacement of the excitonic wavepacket  (MSD$_\text{exc}$, {\bf{e}}). After photo-excitation into the highest-energy eigenstate of the molecule-cavity system (\textit{i.e.}, $|\psi^{1263}\rangle$, which is dominated by the S$_1$ electronic state of HBQ (\textit{i.e.},  $|\beta^{1263}_\text{HBQ}|^2>$ 0.9995, Figure~\ref{fig:transport}{\bf{c}}), with a photon of an energy above the polaritonic manifold, the proton transfers from the hydroxyl oxygen to the nitrogen atom (Figure~\ref{fig:transport}{\bf{a}}). Because this enol to keto transformation is accompanied by a 1.4~eV red-shift of the S$_1$-S$_0$ energy gap, HBQ becomes resonant with the MeB molecules as well as the cavity modes, and enters the dark state manifold around 10~fs after excitation.

% figure proton transfer, WP and MSD

% polariton diffusion

Due to displacements along vibrations parallel to the non-adiabatic coupling vector between this new dark state that is localized on HBQ, on the one hand, and the delocalized bright polaritonic states of the strongly coupled molecule-cavity system on the other hand, population is transferred from HBQ into the bright states~\citep{Groenhof2019,Tichauer2022}. Because these bright polaritonic states have group velocity, the transferred population starts propagating, as shown in Figure~\ref{fig:transport}{\bf{b}}~\citep{Sokolovskii2022}. However, since the population transfer is reversible, this propagation continues ballistically until the population transfers back into the dark state manifold. Due to such reversible exchanges of population between the dark states that lack group velocity and are thus stationary, on the one hand, and the bright polaritonic states with group velocity, on the other hand, the propagation appears as a diffusion process, indicated by the linear increase of the MSD (Figure~\ref{fig:transport}{\bf{b}})~\citep{Sokolovskii2022}. The observation of  diffusion rather than ballistic motion, is in line with experiments on organic micro-cavities~\citep{Rozenman2018}.

% confirm that proton transfer starts transport.

The initial structures for our simulations were sampled from equilibrium QM/MM trajectories at 300~K (SM) and therefore can capture the heterogeneity as indicated by the absorption line-widths of the molecules in Figure~\ref{fig:overview_system}{\bf{e}}. Because of such structural disorder, the ESIPT reaction rates span a distribution. To confirm that the proton transfer in HBQ initiates the polariton-mediated exciton transport process, we show in Figure~S5 (SM) that for a system in which the ESIPT is delayed, also the transport starts at a later point in time, and that this time point coincides with the formation of the HBQ photo-product.

\section{Conclusion}

To summarize, the results of our MD simulations suggest that long-range polariton-mediated exciton transport can be induced with an excited-state proton transfer reaction. While the excitation scheme proposed here resembles the off-resonant laser excitation conditions employed in previous experiments of polariton transport~\citep{Lerario2017, Rozenman2018, Berghuis2022, Balasubrahmaniyam2023}, the absorption cross-section of HBQ should be high enough to initiate the propagation with incoherent light, in particular for a cavity with a thin silver top mirror, which is more than 50\% transparent at the required wave length (Figure~S3, SM).  

\begin{acknowledgement}
We thank Jussi Toppari, Johannes Feist and Ruth Tichauer for fruitful discussions. We thank Arpan Dutta for sharing the absorption spectrum of HBQ in PMMA and Dmitry Morozov for computing the potential energy profiles of proton transfer in HBQ. We thank the Center for Scientific Computing (CSC-IT Center for Science) for very generous computer resources.
\end{acknowledgement}

\begin{funding}
This work was supported by the Academy of Finland (DOI: 10.13039/501100002341, Grants 323996 and 332743).
\end{funding}

\begin{authorcontributions}
G.G. acquired funding; I.S and G.G. performed the Molecular Dynamics simulations. I.S. analysed the data and prepared the figures; G.G. drafted the manuscript, which both authors revised and edited. Both authors have accepted responsibility for the entire content of this manuscript and approved its submission.
\end{authorcontributions}

\begin{conflictofinterest}
The authors state no conflict of interest.
\end{conflictofinterest}

\begin{dataavailabilitystatement}
The datasets generated and/or analyzed during the current study are available from the corresponding author upon reasonable request.
\end{dataavailabilitystatement}

\bibliographystyle{IEEEtran}
\bibliography{bibliography}

\begin{thebibliography}{10}
\providecommand{\url}[1]{#1}
\csname url@samestyle\endcsname
\providecommand{\newblock}{\relax}
\providecommand{\bibinfo}[2]{#2}
\providecommand{\BIBentrySTDinterwordspacing}{\spaceskip=0pt\relax}
\providecommand{\BIBentryALTinterwordstretchfactor}{4}
\providecommand{\BIBentryALTinterwordspacing}{\spaceskip=\fontdimen2\font plus
\BIBentryALTinterwordstretchfactor\fontdimen3\font minus
  \fontdimen4\font\relax}
\providecommand{\BIBforeignlanguage}[2]{{%
\expandafter\ifx\csname l@#1\endcsname\relax
\typeout{** WARNING: IEEEtran.bst: No hyphenation pattern has been}%
\typeout{** loaded for the language `#1'. Using the pattern for}%
\typeout{** the default language instead.}%
\else
\language=\csname l@#1\endcsname
\fi
#2}}
\providecommand{\BIBdecl}{\relax}
\BIBdecl

\bibitem{Mikhnenko2015}
O.~V. Mikhnenko, P.~W.~M. Blom, and T.-Q. Nguyen, ``Exciton diffusion in
  organic semiconductors,'' \emph{Energy Environ. Sci.}, vol.~8, pp.
  1867--1888, 2015.

\bibitem{Pope1977}
J.~B. Aladekomo, S.~Arnold, and M.~Pope, ``Triplet exciton diffusion and double
  photon absorption in tetracene,'' \emph{Phys. Status Solidi}, vol.~80, pp.
  333--340, 1977.

\bibitem{Akselrod2014}
G.~M. Akselrod, P.~B. Deotare, N.~J. Thompson, J.~Lee, W.~A. Tisdale, M.~A. B.
  V.~M. Menon, and V.~Bulovic, ``Visualization of exciton transport in ordered
  and disordered molecular solids,'' \emph{Nat. commun.}, vol.~5, p. 3646,
  2014.

\bibitem{Sneyd2021}
A.~Sneyd, T.~Fukui, D.~Pala\v{c}ek, S.~Prodhan, Y.~Wagner, Y.~Zhang, J.~Sung,
  S.~Collins, T.~Slater, Z.~Andaji-Garmaroudi, L.~MacFarlane,
  J.~Garcia-Hernandez, L.~Wang, G.~Whittell, J.~Hodgkiss, K.~Chen, D.~Beljonne,
  I.~Manners, R.~Friend, and A.~Rao, ``Efficient energy transport in an organic
  semiconductor mediated by transient exciton delocalization,'' \emph{Sci.
  Adv.}, vol.~7, p. eabh4232, 2021.

\bibitem{Kong2022}
F.~F. Kong, X.~J. Tian, Y.~Zhang, Y.~Zhang, G.~Chen, Y.~J. Yu, S.~H. Jing,
  H.~Y. Gao, Y.~Luo, J.~L. Yang, Z.~C. Dong, and J.~G. Hou, ``Wavelike
  electronic energy transfer in donor–acceptor molecular systems through
  quantum coherence,'' \emph{Nat. Nanotechnol.}, vol.~17, pp. 729--736, 2022.

\bibitem{Sneyd2022}
A.~J. Sneyd, D.~Beljonne, and A.~Rao, ``A new frontier in exciton transport:
  Transient delocalization,'' \emph{J. Chem. Phys. Lett.}, vol.~13, pp.
  6820--6830, 2022.

\bibitem{Hildner2023}
S.~Stäter, F.~A. Wenzel, H.~Welz, K.~Kreger, J.~Köhler, H.-W. Schmidt, and
  R.~Hildner, ``Directed gradients in the excited-state energy landscape of
  poly(3-hexylthiophene) nanofibers,'' \emph{J. Am. Chem. Soc.}, vol. 145, pp.
  13\,780--13\,787, 2023.

\bibitem{Lerario2017}
G.~Lerario, D.~Ballarini, A.~Fieramosca, A.~Cannavale, A.~Genco, F.~Mangione,
  S.~Gambino, L.~Dominici, M.~D. Giorgi, G.~Gigli, and D.~Sanvitto,
  ``High-speed flow of interacting organic polaritons,'' \emph{Light Sci.
  Appl.}, vol.~6, p. e16212, 2017.

\bibitem{Rozenman2018}
G.~G. Rozenman, K.~Akulov, A.~Golombek, and T.~Schwartz, ``Long-range transport
  of organic exciton-polaritons revealed by ultrafast microscopy,'' \emph{ACS
  Photonics}, vol.~5, pp. 105--110, 2018.

\bibitem{Forrest2020}
S.~Hou, M.~Khatoniar, K.~Ding, Y.~Qu, A.~Napolov, V.~M. Menon, and S.~R.
  Forrest, ``Ultralong-range energy transport in a disordered organic
  semiconductor at room temperature via coherent exciton-polariton
  propagation,'' \emph{Adv. Mater.}, vol. 32(28), p. 2002127, 2020.

\bibitem{Pandya2021}
R.~Pandya, R.~Y.~S. Chen, Q.~Gu, J.~Sung, C.~Schnedermann, O.~S. Ojambati,
  R.~Chikkaraddy, J.~Gorman, G.~Jacucci, O.~D. Onelli, T.~Willhammar, D.~N.
  Johnstone, S.~M. Collins, P.~A. Midgley, F.~Auras, T.~Baikie, R.~Jayaprakash,
  F.~Mathevet, R.~Soucek, M.~Du, A.~M. Alvertis, A.~Ashoka, S.~Vignolini, D.~G.
  Lidzey, J.~J. Baumberg, R.~H. Friend, T.~Barisien, L.~Legrand, A.~W. Chin,
  J.~Yuen-Zhou, S.~K. Saikin, P.~Kukura, A.~J. Musser, and A.~Rao,
  ``Microcavity-like exciton-polaritons can be the primary photoexcitation in
  bare organic semiconductors,'' \emph{Nat. Commun.}, vol.~12, p. 6519, 2021.

\bibitem{Ostrovskaya2021}
M.~Wurdack, E.~Estrecho, S.~Todd, T.~Yun, M.~Pieczarka, S.~K. Earl, J.~A.
  Davis, C.~Schneider, A.~G. Truscott, and E.~A. Ostrovskaya, ``Motional
  narrowing, ballistic transport, and trapping of room-temperature exciton
  polaritons in an atomically-thin semiconductor,'' \emph{Nat. commun.},
  vol.~12, p. 5366, 2021.

\bibitem{Berghuis2022}
M.~A. Berghuis, R.~H. Tichauer, L.~de~Jong, I.~Sokolovskii, P.~Bai,
  M.~Ramezani, S.~Murai, G.~Groenhof, and J.~G. Rivas, ``Controlling exciton
  propagation in organic crystals through strong coupling to plasmonic
  nanoparticle arrays,'' \emph{ACS Photonics}, vol.~9, p. 123, 2022.

\bibitem{Pandya2022}
R.~Pandya, A.~Ashoka, K.~Georgiou, J.~Sung, R.~Jayaprakash, S.~Renken, L.~Gai,
  Z.~Shen, A.~Rao, and A.~J. Musser, ``Tuning the coherent propagation of
  organic exciton-polaritons through dark state delocalization,'' \emph{Adv.
  Sci.}, vol.~9, p. 2105569, 2022.

\bibitem{Xu2022}
D.~Xu, A.~Mandal, J.~M. Baxter, S.-W. Cheng, I.~Lee, H.~Su, S.~Liu, D.~R.
  Reichman, and M.~Delor, ``Ultrafast imaging of coherent polariton propagation
  and interactions,'' \emph{Nat. commun.}, vol.~14, p. 3881, 2023.

\bibitem{Balasubrahmaniyam2023}
M.~Balasubrahmaniyam, A.~Simkovich, A.~Golombek, G.~Ankonina, and T.~Schwartz,
  ``Unveiling the mixed nature of polaritonic transport: From enhanced
  diffusion to ballistic motion approaching the speed of light,'' \emph{Nat.
  Mater.}, vol.~22, p. 338–344, 2023.

\bibitem{Vahala2003}
K.~J. Vahala, ``Optical microcavities,'' \emph{Nature}, vol. 424, pp. 839--846,
  2003.

\bibitem{Agranovich2007}
V.~M. Agranovich and Y.~N. Gartstein, ``Nature and dynamics of low-energy
  exciton polaritons in semiconductor microcavities,'' \emph{Phys. Rev. B},
  vol.~75, p. 075302, 2007.

\bibitem{Litinskaya2008}
M.~Litinskaya, ``Propagation and localization of polaritons in disordered
  organic microcavities,'' \emph{Phys. Lett. A}, vol. 372, pp. 3898--3903,
  2008.

\bibitem{Feist2015}
J.~Feist and F.~J. Garcia-Vidal, ``Extraordinary exciton conductance induced by
  strong coupling,'' \emph{Phys. Rev. Lett.}, vol. 114, p. 196402, 2015.

\bibitem{Schachenmayer2015}
J.~Schachenmayer, C.~Genes, E.~Tignone, and G.~Pupillo, ``{Cavity enhanced
  transport of excitons},'' \emph{Phys. Rev. Lett.}, vol. 114, p. 196403, 2015.

\bibitem{Engelhardt2023}
G.~Engelhardt and J.~Cao, ``Polariton localization and dispersion properties of
  disordered quantum emitters in multimode microcavities,'' \emph{Phys. Rev.
  Lett.}, vol. 130, p. 213602, 2023.

\bibitem{Aroeira2023}
G.~J.~R. Aroeira, K.~Kairys, and R.~F. Ribeiro, ``Theoretical analysis of
  exciton wave packet dynamics in polaritonic wires,'' \emph{J. Phys. Chem.
  Lett.}, vol.~14, pp. 5681--5691, 2023.

\bibitem{Torma2015}
P.~T{\"{o}}rm{\"{a}} and W.~L. Barnes, ``Strong coupling between surface
  plasmon polaritons and emitters: a review,'' \emph{Rep. Prog. Phys.},
  vol.~78, p. 013901, 2015.

\bibitem{Rider2022}
M.~S. Rider and W.~L. Barnes, ``Something from nothing: linking molecules with
  virtual light,'' \emph{Contemp. Phys.}, vol.~62, no.~4, pp. 217--232, 2022.

\bibitem{Agranovich2003}
V.~M. Agranovich, M.~Litinskaia, and D.~G. Lidzey, ``Cavity polaritons in
  microcavities containing disordered organic semiconductors,'' \emph{Phys.
  Rev. B}, vol.~67, p. 085311, 2003.

\bibitem{Litinskaya2004}
M.~Litinskaya, P.~Reineker, and V.~M. Agranovich, ``Fast polariton relaxation
  in strongly coupled organic microcavities,'' \emph{J. Lumin.}, vol. 110, pp.
  364--372, 2004.

\bibitem{Martinez2019}
L.~A. Mart\'{i}nez-Mart\'{i}nez, E.~Eizner, S.~K\'{e}na-Cohen, and
  J.~Yuen-Zhou, ``Triplet harvesting in the polaritonic regime: A variational
  polaron approach,'' \emph{J. Chem. Phys.}, vol. 151, p. 054106, 2019.

\bibitem{Michetti2008b}
P.~Michetti and G.~C.~L. Rocca, ``Polariton dynamics in disordered
  microcavities,'' \emph{Physica E}, vol.~40, pp. 1926--1929, 2008.

\bibitem{Ribeiro2022}
R.~F. Ribeiro, ``Multimode polariton effects on molecular energy transport and
  spectral fluctuations,'' \emph{Comm. Chem.}, vol.~5, p.~48, 2022.

\bibitem{Freixanet2000}
T.~Freixanet, B.~Sermage, A.~Tiberj, and R.~Planel, ``In-plane propagation of
  excitonic cavity polaritons,'' \emph{Phys. Rev. B}, vol.~61, p. 7233, 2000.

\bibitem{Myers2018}
D.~M. Myers, S.~Mukherjee, J.~Beaumariage, and D.~W. Snoke,
  ``Polariton-enhanced exciton transport,'' \emph{Phys. Rev. B}, vol.~98, p.
  235302, 2018.

\bibitem{Sokolovskii2022}
I.~Sokolovskii, R.~H. Tichauer, D.~Morozov, J.~Feist, and G.~Groenhof,
  ``Multi-scale molecular dynamics simulations of enhanced energy transfer in
  organic molecules under strong coupling,'' \emph{arXiv}, p. 2209.07309, 2022.

\bibitem{Tichauer2023}
R.~H. Tichauer, I.~Sokolovskii, and G.~Groenhof, ``Tuning coherent propagation
  of organic exciton-polaritons through the cavity q-factor,'' \emph{arXiv},
  vol. 2304.13123, 2023.

\bibitem{Wellnitz2022}
D.~Wellnitz, G.~Pupillo, and J.~Schachenmayer, ``Disorder enhanced vibrational
  entanglement and dynamics in polaritonic chemistry,'' \emph{Comm. Phys.},
  vol.~5, p. 120, 2022.

\bibitem{Hutchison2012}
J.~A. Hutchison, T.~Schwartz, C.~Genet, E.~Devaux, and T.~W. Ebbesen,
  ``Modifying chemical landscapes by coupling to vacuum fields,'' \emph{Angew.
  Chem. Int. Ed.}, vol.~51, pp. 1592--1596, 2012.

\bibitem{Kim2009}
C.~H. Kim and T.~Joo, ``Coherent excited state intramolecular proton transfer
  probed by time-resolved fluorescence,'' \emph{Phys. Chem. Chem. Phys.},
  vol.~11, pp. {10\,266--10\,269}, 2009.

\bibitem{Lee2013}
J.~Lee, C.~H. Kim, and T.~Joo, ``Active role of proton in excited state
  intramolecular proton transfer reaction,'' \emph{J. Phys. Chem. A}, vol. 117,
  pp. {1400--1405}, 2013.

\bibitem{Akselrod2013}
G.~M. Akselrod, E.~R. Young, M.~S. Bradley, and V.~Bulovi\'{c}, ``Lasing
  through a strongly-coupled mode by intra-cavity pumping,'' \emph{Opt.
  Express}, vol.~21, pp. 12\,122--12\,128, 2013.

\bibitem{Cargioli2023}
A.~Cargioli, M.~Lednev, L.~Lavista, A.~Camposeo, A.~Sassella, D.~Pisignano,
  A.~Tredicucci, F.~J. Garcia-Vidal, J.~Feist, and L.~Persano, ``Active control
  of polariton-enabled long-range energy transfer,'' \emph{arXiv}, p.
  2310.04121, 2023.

\bibitem{Luk2017}
H.~L. Luk, J.~Feist, J.~J. Toppari, and G.~Groenhof, ``Multiscale molecular
  dynamics simulations of polaritonic chemistry,'' \emph{J. Chem. Theory
  Comput.}, vol.~13, pp. 4324--4335, 2017.

\bibitem{Tichauer2021}
R.~H. Tichauer, J.~Feist, and G.~Groenhof, ``Multi-scale dynamics simulations
  of molecular polaritons: the effect of multiple cavity modes on polariton
  relaxation,'' \emph{J. Chem. Phys.}, vol. 154, p. 104112, 2021.

\bibitem{delPino2015}
J.~{del Pino}, J.~Feist, and F.~J. {Garcia-Vidal}, ``{Quantum Theory of
  Collective Strong Coupling of Molecular Vibrations with a Microcavity
  Mode},'' \emph{New J. Phys.}, vol.~17, no.~5, p. 053040, May 2015.

\bibitem{Eizner2019}
E.~Eizner, L.~A. Mart\'{i}nez-Mart\'{i}nez, J.~{Yuen-Shou}, and
  S.~K\'{e}na-Cohen, ``Inverting singlet and triplet excited states using
  strong light-matter coupling,'' \emph{Sci. Adv.}, vol.~5, p. aax4482, 2019.

\bibitem{Hohenberg1964}
P.~Hohenberg and W.~Kohn, ``Imhomogeneous electron gas,'' \emph{Phys. Rev.},
  vol. 136, pp. 864--871, 1964.

\bibitem{Runge1984}
E.~Runge and E.~K.~U. Gross, ``Density-functional theory for time-dependent
  systems,'' \emph{Phys. Rev. Lett}, vol.~52, pp. 997--1000, 1984.

\bibitem{Hirata1999}
S.~Hirata and M.~Head-Gordon, ``Time-dependent density functional theory within
  the tamm–dancoff approximation,'' \emph{Chem. Phys. Lett.}, vol. 314, pp.
  291--299, 1999.

\bibitem{Becke1993}
A.~D. Becke, ``{A new mixing of Hartree-Fock and local density-functional
  theories},'' \emph{J. Chem. Phys.}, vol.~98, p. 1372, 1993.

\bibitem{Yanai2004}
T.~Yanai, D.~P. Tew, and N.~C. Handy, ``A new hybrid exchange-correlation
  functional using the coulomb-attenuating method (cam-b3lyp),'' \emph{Chem.
  Phys. Lett.}, vol. 393, pp. 51--57, 2004.

\bibitem{Ditchfield1971}
R.~Ditchfield, W.~J. Hehre, and J.~A. Pople, ``Self-consistent
  molecular-orbital methods. ix. an extended gaussian-type basis for
  molecular-orbital studies of organic molecules,'' \emph{J. Chem. Phys.},
  vol.~54, pp. 724--728, 1971.

\bibitem{Horta2016}
B.~A.~C. Horta, P.~T. Merz, P.~F.~J. Fuchs, J.~Dolenc, S.~Riniker, and P.~H.
  Hünenberger, ``A gromos-compatible force field for small organic molecules
  in the condensed phase: The 2016h66 parameter set,'' \emph{J. Chem. Theory.
  Comput.}, vol.~12, pp. 3825--3850, 2016.

\bibitem{Staroverov2003}
V.~N. Staroverov, G.~E. Scuseria, J.~Tao, and J.~P. Perdew, ``Comparative
  assessment of a new nonempirical density functional: Molecules and
  hydrogen-bonded complexes,'' \emph{J. Chem. Phys.}, vol. 119, no.~23, pp.
  12\,129--12\,137, 2003.

\bibitem{Picconi2021}
D.~Picconi, ``Nonadiabatic quantum dynamics of the coherent excited state
  intramolecular proton transfer of 10-hydroxybenzo [h] quinoline,''
  \emph{Photochem. Photobiol. Sci.}, vol.~20, pp. 1455--1473, 2021.

\bibitem{Casida1998}
M.~E. Casida, C.~Jamorski, K.~C. Casida, and D.~R. Salahub, ``Molecular
  excitation energies to high-lying bound states from time-dependent
  density-functional response theory: Characterization and correction of the
  time-dependent local density approximation ionization threshold,'' \emph{J.
  Chem. Phys.}, vol. 108, pp. 4439--4449, 1998.

\bibitem{Becke97}
A.~D. Becke, ``Density-functional thermochemistry. v. systematic optimization
  of exchange-correlation functionals,'' \emph{J. Chem. Phys.}, vol. 107, pp.
  8554--8560, 1997.

\bibitem{Jorgensen1983}
W.~L. Jorgensen, J.~Chandrasekhar, J.~D. Madura, R.~W. Impey, and M.~L. Klein,
  ``Comparison of simple potential functions for simulatin liquid water,''
  \emph{J. Chem. Phys.}, vol.~79, pp. 926--935, 1983.

\bibitem{Michetti2005}
P.~Michetti and G.~C.~L. Rocca, ``Polariton states in disordered organic
  microcavities,'' \emph{Phys. Rev. B.}, vol.~71, p. 115320, 2005.

\bibitem{Schwartz2013}
T.~Schwartz, J.~A. Hutchison, J.~Leonard, C.~Genet, S.~Haacke, and T.~W.
  Ebbesen, ``Polariton dynamics under strong light-molecule coupling,''
  \emph{ChemPhysChem}, vol.~14, pp. 125--131, 2013.

\bibitem{Ulusoy2020}
I.~S. Ulusoy and O.~Vendrell, ``Dynamics and spectroscopy of molecular
  ensembles in a lossy microcavity,'' \emph{J. Chem. Phys.}, vol. 153, p.
  044108, 2020.

\bibitem{Antoniou2020}
P.~Antoniou, F.~Suchanek, J.~F. Varner, and J.~J. {Foley IV}, ``Role of cavity
  losses on nonadiabatic couplings and dynamics in polaritonic chemistry,''
  \emph{J. Phys. Chem. Lett.}, vol.~11, pp. 9063--9069, 2020.

\bibitem{Felicetti2020}
S.~Felicetti, J.~Fregoni, T.~Schnappinger, S.~Reiter, R.~de~Vivie-Riedle, and
  J.~Feist, ``Photoprotecting uracil by coupling with lossy nanocavities,''
  \emph{J. Chem. Phys. Lett.}, vol.~11, pp. 8810--8818, 2020.

\bibitem{Hu2022}
D.~Hu, A.~Mandal, B.~M. Weight, and P.~Huo, ``Quasi-diabatic propagation scheme
  for simulating polariton chemistry,'' \emph{J. Chem. Phys.}, vol. 157, p.
  194109, 2022.

\bibitem{Groenhof2019}
G.~Groenhof, C.~Climent, J.~Feist, D.~Morozov, and J.~J. Toppari, ``Tracking
  polariton relaxation with multiscale molecular dynamics simulations,''
  \emph{J. Chem. Phys. Lett.}, vol.~10, pp. 5476--5483, 2019.

\bibitem{Tichauer2022}
R.~H. Tichauer, D.~Morozov, I.~Sokolovskii, J.~J. Toppari, and G.~Groenhof,
  ``Identifying vibrations that control non-adiabatic relaxation of polaritons
  in strongly coupled molecule-cavity systems,'' \emph{J. Phys. Chem. Lett.},
  vol.~13, pp. 6259--6267, 2022.

\end{thebibliography}


\providecommand{\latin}[1]{#1}
\makeatletter
\providecommand{\doi}
  {\begingroup\let\do\@makeother\dospecials
  \catcode`\{=1 \catcode`\}=2 \doi@aux}
\providecommand{\doi@aux}[1]{\endgroup\texttt{#1}}
\makeatother
\providecommand*\mcitethebibliography{\thebibliography}
\csname @ifundefined\endcsname{endmcitethebibliography}
  {\let\endmcitethebibliography\endthebibliography}{}
\begin{mcitethebibliography}{44}
\providecommand*\natexlab[1]{#1}
\providecommand*\mciteSetBstSublistMode[1]{}
\providecommand*\mciteSetBstMaxWidthForm[2]{}
\providecommand*\mciteBstWouldAddEndPuncttrue
  {\def\EndOfBibitem{\unskip.}}
\providecommand*\mciteBstWouldAddEndPunctfalse
  {\let\EndOfBibitem\relax}
\providecommand*\mciteSetBstMidEndSepPunct[3]{}
\providecommand*\mciteSetBstSublistLabelBeginEnd[3]{}
\providecommand*\EndOfBibitem{}
\mciteSetBstSublistMode{f}
\mciteSetBstMaxWidthForm{subitem}{(\alph{mcitesubitemcount})}
\mciteSetBstSublistLabelBeginEnd
  {\mcitemaxwidthsubitemform\space}
  {\relax}
  {\relax}

\bibitem[Forn-D\'{i}az \latin{et~al.}(2019)Forn-D\'{i}az, Lamata, Rico, Kono,
  and Solano]{Forn-Diaz2019}
Forn-D\'{i}az,~P.; Lamata,~L.; Rico,~E.; Kono,~J.; Solano,~E. Ultrastrong
  coupling regimes of light-matter interaction. \emph{Rev. Mod. Phys.}
  \textbf{2019}, \emph{91}, 025005\relax
\mciteBstWouldAddEndPuncttrue
\mciteSetBstMidEndSepPunct{\mcitedefaultmidpunct}
{\mcitedefaultendpunct}{\mcitedefaultseppunct}\relax
\EndOfBibitem
\bibitem[Jaynes and Cummings(1963)Jaynes, and Cummings]{Jaynes1963}
Jaynes,~E.~T.; Cummings,~F.~W. Comparison of quantum and semiclassical
  radiation theories with application to the beam maser. \emph{Proc. IEEE}
  \textbf{1963}, \emph{51}, 89--109\relax
\mciteBstWouldAddEndPuncttrue
\mciteSetBstMidEndSepPunct{\mcitedefaultmidpunct}
{\mcitedefaultendpunct}{\mcitedefaultseppunct}\relax
\EndOfBibitem
\bibitem[Tavis and Cummings(1969)Tavis, and Cummings]{Tavis1969}
Tavis,~M.; Cummings,~F.~W. Approximate solutions for an N-molecule
  radiation-field Hamiltonian. \emph{Phys. Rev.} \textbf{1969}, \emph{188},
  692--695\relax
\mciteBstWouldAddEndPuncttrue
\mciteSetBstMidEndSepPunct{\mcitedefaultmidpunct}
{\mcitedefaultendpunct}{\mcitedefaultseppunct}\relax
\EndOfBibitem
\bibitem[Michetti and Rocca(2005)Michetti, and Rocca]{Michetti2005}
Michetti,~P.; Rocca,~G. C.~L. Polariton states in disordered organic
  microcavities. \emph{Phys. Rev. B.} \textbf{2005}, \emph{71}, 115320\relax
\mciteBstWouldAddEndPuncttrue
\mciteSetBstMidEndSepPunct{\mcitedefaultmidpunct}
{\mcitedefaultendpunct}{\mcitedefaultseppunct}\relax
\EndOfBibitem
\bibitem[Tichauer \latin{et~al.}(2021)Tichauer, Feist, and
  Groenhof]{Tichauer2021}
Tichauer,~R.~H.; Feist,~J.; Groenhof,~G. Multi-scale Dynamics Simulations of
  Molecular Polaritons: the Effect of Multiple Cavity Modes on Polariton
  Relaxation. \emph{J. Chem. Phys.} \textbf{2021}, \emph{154}, 104112\relax
\mciteBstWouldAddEndPuncttrue
\mciteSetBstMidEndSepPunct{\mcitedefaultmidpunct}
{\mcitedefaultendpunct}{\mcitedefaultseppunct}\relax
\EndOfBibitem
\bibitem[Warshel and Levitt(1976)Warshel, and Levitt]{Warshel1976b}
Warshel,~A.; Levitt,~M. Theoretical studies of enzymatic reactions: Dielectric,
  electrostatic and steric stabilization of carbonium ion in the reaction of
  lysozyme. \emph{J. Mol. Biol.} \textbf{1976}, \emph{103}, 227--249\relax
\mciteBstWouldAddEndPuncttrue
\mciteSetBstMidEndSepPunct{\mcitedefaultmidpunct}
{\mcitedefaultendpunct}{\mcitedefaultseppunct}\relax
\EndOfBibitem
\bibitem[Boggio-Pasqua \latin{et~al.}(2012)Boggio-Pasqua, Burmeister, Robb, and
  Groenhof]{Boggio-Pasqua2012}
Boggio-Pasqua,~M.; Burmeister,~C.~F.; Robb,~M.~A.; Groenhof,~G. Photochemical
  reactions in biological systems: probing the effect of the environment by
  means of hybrid quantum chemistry/molecular mechanics simulations.
  \emph{Phys. Chem. Chem. Phys.} \textbf{2012}, \emph{14}, 7912--7928\relax
\mciteBstWouldAddEndPuncttrue
\mciteSetBstMidEndSepPunct{\mcitedefaultmidpunct}
{\mcitedefaultendpunct}{\mcitedefaultseppunct}\relax
\EndOfBibitem
\bibitem[Ehrenfest(1927)]{Ehrenfest1927}
Ehrenfest,~P. Bemerkung über die angen\"{a}herte G\"{u}ltigkeit der
  klassischen Mechanik innerhalb der Quantenmechanik. \emph{Z. Phys.}
  \textbf{1927}, \emph{45}, 445--457\relax
\mciteBstWouldAddEndPuncttrue
\mciteSetBstMidEndSepPunct{\mcitedefaultmidpunct}
{\mcitedefaultendpunct}{\mcitedefaultseppunct}\relax
\EndOfBibitem
\bibitem[Felicetti \latin{et~al.}(2020)Felicetti, Fregoni, Schnappinger,
  Reiter, de~Vivie-Riedle, and Feist]{Felicetti2020}
Felicetti,~S.; Fregoni,~J.; Schnappinger,~T.; Reiter,~S.; de~Vivie-Riedle,~R.;
  Feist,~J. Photoprotecting Uracil by Coupling with Lossy Nanocavities.
  \emph{J. Chem. Phys. Lett.} \textbf{2020}, \emph{11}, 8810--8818\relax
\mciteBstWouldAddEndPuncttrue
\mciteSetBstMidEndSepPunct{\mcitedefaultmidpunct}
{\mcitedefaultendpunct}{\mcitedefaultseppunct}\relax
\EndOfBibitem
\bibitem[Ulusoy and Vendrell(2020)Ulusoy, and Vendrell]{Ulusoy2020}
Ulusoy,~I.~S.; Vendrell,~O. Dynamics and spectroscopy of molecular ensembles in
  a lossy microcavity. \emph{J. Chem. Phys.} \textbf{2020}, \emph{153},
  044108\relax
\mciteBstWouldAddEndPuncttrue
\mciteSetBstMidEndSepPunct{\mcitedefaultmidpunct}
{\mcitedefaultendpunct}{\mcitedefaultseppunct}\relax
\EndOfBibitem
\bibitem[Antoniou \latin{et~al.}(2020)Antoniou, Suchanek, Varner, and {Foley
  IV}]{Antoniou2020}
Antoniou,~P.; Suchanek,~F.; Varner,~J.~F.; {Foley IV},~J.~J. Role of Cavity
  Losses on Nonadiabatic Couplings and Dynamics in Polaritonic Chemistry.
  \emph{J. Phys. Chem. Lett.} \textbf{2020}, \emph{11}, 9063--9069\relax
\mciteBstWouldAddEndPuncttrue
\mciteSetBstMidEndSepPunct{\mcitedefaultmidpunct}
{\mcitedefaultendpunct}{\mcitedefaultseppunct}\relax
\EndOfBibitem
\bibitem[Hu \latin{et~al.}(2022)Hu, Mandal, Weight, and Huo]{Hu2022}
Hu,~D.; Mandal,~A.; Weight,~B.~M.; Huo,~P. Quasi-diabatic propagation scheme
  for simulating polariton chemistry. \emph{J. Chem. Phys.} \textbf{2022},
  \emph{157}, 194109\relax
\mciteBstWouldAddEndPuncttrue
\mciteSetBstMidEndSepPunct{\mcitedefaultmidpunct}
{\mcitedefaultendpunct}{\mcitedefaultseppunct}\relax
\EndOfBibitem
\bibitem[Horta \latin{et~al.}(2016)Horta, Merz, Fuchs, Dolenc, Riniker, and
  Hünenberger]{Horta2016}
Horta,~B. A.~C.; Merz,~P.~T.; Fuchs,~P. F.~J.; Dolenc,~J.; Riniker,~S.;
  Hünenberger,~P.~H. A GROMOS-Compatible Force Field for Small Organic
  Molecules in the Condensed Phase: The 2016H66 Parameter Set. \emph{J. Chem.
  Theory. Comput.} \textbf{2016}, \emph{12}, 3825--3850\relax
\mciteBstWouldAddEndPuncttrue
\mciteSetBstMidEndSepPunct{\mcitedefaultmidpunct}
{\mcitedefaultendpunct}{\mcitedefaultseppunct}\relax
\EndOfBibitem
\bibitem[Hess \latin{et~al.}(1997)Hess, Bekker, Berendsen, and
  Fraaije]{Hess1997}
Hess,~B.; Bekker,~H.; Berendsen,~H. J.~C.; Fraaije,~J. G. E.~M. {LINCS: A
  linear constraint solver for molecular simulations}. \emph{J. Comput. Chem.}
  \textbf{1997}, \emph{18}, 1463--1472\relax
\mciteBstWouldAddEndPuncttrue
\mciteSetBstMidEndSepPunct{\mcitedefaultmidpunct}
{\mcitedefaultendpunct}{\mcitedefaultseppunct}\relax
\EndOfBibitem
\bibitem[Berendsen \latin{et~al.}(1984)Berendsen, Postma, van Gunsteren, la,
  and Haak]{Berendsen1984}
Berendsen,~H.; Postma,~J.; van Gunsteren,~W.; la,~A.~D.; Haak,~J. Molecular
  dynamics with coupling to an external bath. \emph{J. Chem. Phys.}
  \textbf{1984}, \emph{81}, 3684--3690\relax
\mciteBstWouldAddEndPuncttrue
\mciteSetBstMidEndSepPunct{\mcitedefaultmidpunct}
{\mcitedefaultendpunct}{\mcitedefaultseppunct}\relax
\EndOfBibitem
\bibitem[Becke(1993)]{Becke1993}
Becke,~A.~D. {A new mixing of Hartree-Fock and local density-functional
  theories}. \emph{J. Chem. Phys.} \textbf{1993}, \emph{98}, 1372\relax
\mciteBstWouldAddEndPuncttrue
\mciteSetBstMidEndSepPunct{\mcitedefaultmidpunct}
{\mcitedefaultendpunct}{\mcitedefaultseppunct}\relax
\EndOfBibitem
\bibitem[Lee \latin{et~al.}(1988)Lee, Yang, and Parr]{Lee1988}
Lee,~C.~T.; Yang,~W.~T.; Parr,~R.~G. Development of the Colle-Salvet
  correlation-energy formula into a functional of the electron density.
  \emph{Phys. Rev. B} \textbf{1988}, \emph{37}, 785--789\relax
\mciteBstWouldAddEndPuncttrue
\mciteSetBstMidEndSepPunct{\mcitedefaultmidpunct}
{\mcitedefaultendpunct}{\mcitedefaultseppunct}\relax
\EndOfBibitem
\bibitem[Yanai \latin{et~al.}(2004)Yanai, Tew, and Handy]{Yanai2004}
Yanai,~T.; Tew,~D.~P.; Handy,~N.~C. A new hybrid exchange-correlation
  functional using the Coulomb-attenuating method (CAM-B3LYP). \emph{Chem.
  Phys. Lett.} \textbf{2004}, \emph{393}, 51--57\relax
\mciteBstWouldAddEndPuncttrue
\mciteSetBstMidEndSepPunct{\mcitedefaultmidpunct}
{\mcitedefaultendpunct}{\mcitedefaultseppunct}\relax
\EndOfBibitem
\bibitem[Dunning(1970)]{Dunning1970}
Dunning,~T.~H. Basis Functions for Use in Molecular Calculations. I.
  Contractions of (9s5p) Atomic Basis Sets for the First-Row Atoms. \emph{J.
  Chem. Phys.} \textbf{1970}, \emph{53}, 2823--2833\relax
\mciteBstWouldAddEndPuncttrue
\mciteSetBstMidEndSepPunct{\mcitedefaultmidpunct}
{\mcitedefaultendpunct}{\mcitedefaultseppunct}\relax
\EndOfBibitem
\bibitem[Runge and Gross(1984)Runge, and Gross]{Runge1984}
Runge,~E.; Gross,~E. K.~U. Density-Functional Theory for Time-Dependent
  Systems. \emph{Phys. Rev. Lett} \textbf{1984}, \emph{52}, 997--1000\relax
\mciteBstWouldAddEndPuncttrue
\mciteSetBstMidEndSepPunct{\mcitedefaultmidpunct}
{\mcitedefaultendpunct}{\mcitedefaultseppunct}\relax
\EndOfBibitem
\bibitem[Hirata and Head-Gordon(1999)Hirata, and Head-Gordon]{Hirata1999}
Hirata,~S.; Head-Gordon,~M. Time-dependent density functional theory within the
  Tamm–Dancoff approximation. \emph{Chem. Phys. Lett.} \textbf{1999},
  \emph{314}, 291--299\relax
\mciteBstWouldAddEndPuncttrue
\mciteSetBstMidEndSepPunct{\mcitedefaultmidpunct}
{\mcitedefaultendpunct}{\mcitedefaultseppunct}\relax
\EndOfBibitem
\bibitem[Hess \latin{et~al.}(2008)Hess, Kutzner, van~der Spoel, and
  Lindahl]{Hess2008}
Hess,~B.; Kutzner,~C.; van~der Spoel,~D.; Lindahl,~E. GROMACS 4: Algorithms for
  Highly Efficient, Load-Balanced, and Scalable Molecular Simulation. \emph{J.
  Chem. Theory Comput.} \textbf{2008}, \emph{4}, 435--447\relax
\mciteBstWouldAddEndPuncttrue
\mciteSetBstMidEndSepPunct{\mcitedefaultmidpunct}
{\mcitedefaultendpunct}{\mcitedefaultseppunct}\relax
\EndOfBibitem
\bibitem[Ufimtsev and Mart\'{i}nez(2009)Ufimtsev, and
  Mart\'{i}nez]{Ufimtsev2009}
Ufimtsev,~I.~S.; Mart\'{i}nez,~T.~J. Quantum Chemistry on Graphical Processing
  Units. 3. Analytical Energy Gradients and First Principles Molecular
  Dynamics. \emph{J. Chem. Theory Comput.} \textbf{2009}, \emph{5},
  2619--2628\relax
\mciteBstWouldAddEndPuncttrue
\mciteSetBstMidEndSepPunct{\mcitedefaultmidpunct}
{\mcitedefaultendpunct}{\mcitedefaultseppunct}\relax
\EndOfBibitem
\bibitem[Titov \latin{et~al.}(2013)Titov, Ufimtsev, Luehr, and
  Mart\'{i}nez]{Titov2013}
Titov,~A.~V.; Ufimtsev,~I.~S.; Luehr,~N.; Mart\'{i}nez,~T.~J. Generating
  Efficient Quantum Chemistry Codes for Novel Architectures. \emph{J. Chem.
  Theory Comput.} \textbf{2013}, \emph{9}, 213--221\relax
\mciteBstWouldAddEndPuncttrue
\mciteSetBstMidEndSepPunct{\mcitedefaultmidpunct}
{\mcitedefaultendpunct}{\mcitedefaultseppunct}\relax
\EndOfBibitem
\bibitem[Jorgensen \latin{et~al.}(1983)Jorgensen, Chandrasekhar, Madura, Impey,
  and Klein]{Jorgensen1983}
Jorgensen,~W.~L.; Chandrasekhar,~J.; Madura,~J.~D.; Impey,~R.~W.; Klein,~M.~L.
  Comparison of simple potential functions for simulatin liquid water. \emph{J.
  Chem. Phys.} \textbf{1983}, \emph{79}, 926--935\relax
\mciteBstWouldAddEndPuncttrue
\mciteSetBstMidEndSepPunct{\mcitedefaultmidpunct}
{\mcitedefaultendpunct}{\mcitedefaultseppunct}\relax
\EndOfBibitem
\bibitem[Duan \latin{et~al.}({2003})Duan, Wu, Chowdhury, Lee, Xiong, Zhang,
  Yang, Cieplak, Luo, Lee, Caldwell, Wang, and Kollman]{Duan2003}
Duan,~Y.; Wu,~C.; Chowdhury,~S.; Lee,~M.~C.; Xiong,~G.~M.; Zhang,~W.; Yang,~R.;
  Cieplak,~P.; Luo,~R.; Lee,~T.; Caldwell,~J.; Wang,~J.~M.; Kollman,~P. {A
  point-charge force field for molecular mechanics simulations of proteins
  based on condensed-phase quantum mechanical calculations}. \emph{J. Comput.
  Chem.} \textbf{{2003}}, \emph{{24}}, {1999--2012}\relax
\mciteBstWouldAddEndPuncttrue
\mciteSetBstMidEndSepPunct{\mcitedefaultmidpunct}
{\mcitedefaultendpunct}{\mcitedefaultseppunct}\relax
\EndOfBibitem
\bibitem[Bayly \latin{et~al.}({1993})Bayly, Cieplak, Cornell, and
  Kollman]{Bayly1993}
Bayly,~C.~I.; Cieplak,~P.; Cornell,~W.~D.; Kollman,~P.~A. {A well-behaved
  electrostatic potential bsed method using charge restraints for deriving
  atomic charges - the RESP model}. \emph{{J. Phys. Chem.}} \textbf{{1993}},
  \emph{{97}}, {10269--10280}\relax
\mciteBstWouldAddEndPuncttrue
\mciteSetBstMidEndSepPunct{\mcitedefaultmidpunct}
{\mcitedefaultendpunct}{\mcitedefaultseppunct}\relax
\EndOfBibitem
\bibitem[Tomasi \latin{et~al.}(2005)Tomasi, Mennucci, and Cammi]{Tomasi2005}
Tomasi,~J.; Mennucci,~B.; Cammi,~R. Quantum Mechanical Continuum Solvation
  Models. \emph{Chem. Rev.} \textbf{2005}, \emph{105}, 2999--3094\relax
\mciteBstWouldAddEndPuncttrue
\mciteSetBstMidEndSepPunct{\mcitedefaultmidpunct}
{\mcitedefaultendpunct}{\mcitedefaultseppunct}\relax
\EndOfBibitem
\bibitem[Essmann \latin{et~al.}(1995)Essmann, Perera, Berkowitz, Darden, Lee,
  and Pedersen]{Essmann1995}
Essmann,~U.; Perera,~L.; Berkowitz,~M.~L.; Darden,~T.; Lee,~H.; Pedersen,~L.~G.
  A smooth particle mesh {E}wald potential. \emph{J. Chem. Phys} \textbf{1995},
  \emph{103}, 8577--8592\relax
\mciteBstWouldAddEndPuncttrue
\mciteSetBstMidEndSepPunct{\mcitedefaultmidpunct}
{\mcitedefaultendpunct}{\mcitedefaultseppunct}\relax
\EndOfBibitem
\bibitem[Bussi \latin{et~al.}(2007)Bussi, Donadio, and Parrinello]{Bussi2007}
Bussi,~G.; Donadio,~D.; Parrinello,~M. Canonical sampling through velocity
  rescaling. \emph{J. Chem. Phys.} \textbf{2007}, \emph{126}, 014101\relax
\mciteBstWouldAddEndPuncttrue
\mciteSetBstMidEndSepPunct{\mcitedefaultmidpunct}
{\mcitedefaultendpunct}{\mcitedefaultseppunct}\relax
\EndOfBibitem
\bibitem[Miyamoto and Kollman(1992)Miyamoto, and Kollman]{Miyamoto1992}
Miyamoto,~S.; Kollman,~P.~A. {SETTLE}: An analytical version of the {SHAKE} and
  {RATTLE} algorithms for rigid water molecules. \emph{J. Comp. Chem.}
  \textbf{1992}, \emph{18}, 1463--1472\relax
\mciteBstWouldAddEndPuncttrue
\mciteSetBstMidEndSepPunct{\mcitedefaultmidpunct}
{\mcitedefaultendpunct}{\mcitedefaultseppunct}\relax
\EndOfBibitem
\bibitem[Becke(1997)]{Becke97}
Becke,~A.~D. Density-functional thermochemistry. V. Systematic optimization of
  exchange-correlation functionals. \emph{J. Chem. Phys.} \textbf{1997},
  \emph{107}, 8554--8560\relax
\mciteBstWouldAddEndPuncttrue
\mciteSetBstMidEndSepPunct{\mcitedefaultmidpunct}
{\mcitedefaultendpunct}{\mcitedefaultseppunct}\relax
\EndOfBibitem
\bibitem[Schwartz \latin{et~al.}(2013)Schwartz, Hutchison, Leonard, Genet,
  Haacke, and Ebbesen]{Schwartz2013}
Schwartz,~T.; Hutchison,~J.~A.; Leonard,~J.; Genet,~C.; Haacke,~S.;
  Ebbesen,~T.~W. Polariton Dynamics under Strong Light-Molecule Coupling.
  \emph{ChemPhysChem} \textbf{2013}, \emph{14}, 125--131\relax
\mciteBstWouldAddEndPuncttrue
\mciteSetBstMidEndSepPunct{\mcitedefaultmidpunct}
{\mcitedefaultendpunct}{\mcitedefaultseppunct}\relax
\EndOfBibitem
\bibitem[George \latin{et~al.}(2015)George, Wang, Chervy, Canaguier-Durand,
  Schaeffer, Lehn, Hutchison, Genet, and Ebbesen]{George2015}
George,~J.; Wang,~S.; Chervy,~T.; Canaguier-Durand,~A.; Schaeffer,~G.;
  Lehn,~J.-M.; Hutchison,~J.~A.; Genet,~C.; Ebbesen,~T.~W. Ultra-strong
  coupling of molecular materials: spectroscopy and dynamics. \emph{Faraday
  Discuss.} \textbf{2015}, \emph{178}, 281--294\relax
\mciteBstWouldAddEndPuncttrue
\mciteSetBstMidEndSepPunct{\mcitedefaultmidpunct}
{\mcitedefaultendpunct}{\mcitedefaultseppunct}\relax
\EndOfBibitem
\bibitem[Rozenman \latin{et~al.}(2018)Rozenman, Akulov, Golombek, and
  Schwartz]{Rozenman2018}
Rozenman,~G.~G.; Akulov,~K.; Golombek,~A.; Schwartz,~T. Long-Range Transport of
  Organic Exciton-Polaritons Revealed by Ultrafast Microscopy. \emph{ACS
  Photonics} \textbf{2018}, \emph{5}, 105--110\relax
\mciteBstWouldAddEndPuncttrue
\mciteSetBstMidEndSepPunct{\mcitedefaultmidpunct}
{\mcitedefaultendpunct}{\mcitedefaultseppunct}\relax
\EndOfBibitem
\bibitem[Granucci \latin{et~al.}(2001)Granucci, Persico, and
  Toniolo]{Granucci2001}
Granucci,~G.; Persico,~M.; Toniolo,~A. Direct semiclassical simulation of
  photochemical processes with semiempirical wave functions. \emph{J. Chem.
  Phys.} \textbf{2001}, \emph{114}, 10608--10615\relax
\mciteBstWouldAddEndPuncttrue
\mciteSetBstMidEndSepPunct{\mcitedefaultmidpunct}
{\mcitedefaultendpunct}{\mcitedefaultseppunct}\relax
\EndOfBibitem
\bibitem[Frisch \latin{et~al.}(2016)Frisch, Trucks, Schlegel, Scuseria, Robb,
  Cheeseman, Scalmani, Barone, Petersson, Nakatsuji, Li, Caricato, Marenich,
  Bloino, Janesko, Gomperts, Mennucci, Hratchian, Ortiz, Izmaylov, Sonnenberg,
  Williams-Young, Ding, Lipparini, Egidi, Goings, Peng, Petrone, Henderson,
  Ranasinghe, Zakrzewski, Gao, Rega, Zheng, Liang, Hada, Ehara, Toyota, Fukuda,
  Hasegawa, Ishida, Nakajima, Honda, Kitao, Nakai, Vreven, Throssell,
  Montgomery, Peralta, Ogliaro, Bearpark, Heyd, Brothers, Kudin, Staroverov,
  Keith, Kobayashi, Normand, Raghavachari, Rendell, Burant, Iyengar, Tomasi,
  Cossi, Millam, Klene, Adamo, Cammi, Ochterski, Martin, Morokuma, Farkas,
  Foresman, and Fox]{g16}
Frisch,~M.~J.; Trucks,~G.~W.; Schlegel,~H.~B.; Scuseria,~G.~E.; Robb,~M.~A.;
  Cheeseman,~J.~R.; Scalmani,~G.; Barone,~V.; Petersson,~G.~A.; Nakatsuji,~H.;
  Li,~X.; Caricato,~M.; Marenich,~A.~V.; Bloino,~J.; Janesko,~B.~G.;
  Gomperts,~R.; Mennucci,~B.; Hratchian,~H.~P.; Ortiz,~J.~V.; Izmaylov,~A.~F.;
  Sonnenberg,~J.~L.; Williams-Young,~D.; Ding,~F.; Lipparini,~F.; Egidi,~F.;
  Goings,~J.; Peng,~B.; Petrone,~A.; Henderson,~T.; Ranasinghe,~D.;
  Zakrzewski,~V.~G.; Gao,~J.; Rega,~N.; Zheng,~G.; Liang,~W.; Hada,~M.;
  Ehara,~M.; Toyota,~K.; Fukuda,~R.; Hasegawa,~J.; Ishida,~M.; Nakajima,~T.;
  Honda,~Y.; Kitao,~O.; Nakai,~H.; Vreven,~T.; Throssell,~K.;
  Montgomery,~J.~A.,~{Jr.}; Peralta,~J.~E.; Ogliaro,~F.; Bearpark,~M.~J.;
  Heyd,~J.~J.; Brothers,~E.~N.; Kudin,~K.~N.; Staroverov,~V.~N.; Keith,~T.~A.;
  Kobayashi,~R.; Normand,~J.; Raghavachari,~K.; Rendell,~A.~P.; Burant,~J.~C.;
  Iyengar,~S.~S.; Tomasi,~J.; Cossi,~M.; Millam,~J.~M.; Klene,~M.; Adamo,~C.;
  Cammi,~R.; Ochterski,~J.~W.; Martin,~R.~L.; Morokuma,~K.; Farkas,~O.;
  Foresman,~J.~B.; Fox,~D.~J. Gaussian˜16 {R}evision {C}.01. 2016; Gaussian
  Inc. Wallingford CT\relax
\mciteBstWouldAddEndPuncttrue
\mciteSetBstMidEndSepPunct{\mcitedefaultmidpunct}
{\mcitedefaultendpunct}{\mcitedefaultseppunct}\relax
\EndOfBibitem
\bibitem[Modi \latin{et~al.}(2019)Modi, Donnini, Groenhof, and
  Morozov]{Modi2019}
Modi,~V.; Donnini,~S.; Groenhof,~G.; Morozov,~D. Protonation of the Biliverdin
  IX$\alpha$ Chromophore in the Red and Far- Red Photoactive States of a
  Bacteriophytochrome. \emph{J. Phys. Chem. B} \textbf{2019}, \emph{123},
  2325--2334\relax
\mciteBstWouldAddEndPuncttrue
\mciteSetBstMidEndSepPunct{\mcitedefaultmidpunct}
{\mcitedefaultendpunct}{\mcitedefaultseppunct}\relax
\EndOfBibitem
\bibitem[Agranovich and Gartstein(2007)Agranovich, and
  Gartstein]{Agranovich2007}
Agranovich,~V.~M.; Gartstein,~Y.~N. Nature and Dynamics of Low-Energy Exciton
  Polaritons in Semiconductor Microcavities. \emph{Phys. Rev. B} \textbf{2007},
  \emph{75}, 075302\relax
\mciteBstWouldAddEndPuncttrue
\mciteSetBstMidEndSepPunct{\mcitedefaultmidpunct}
{\mcitedefaultendpunct}{\mcitedefaultseppunct}\relax
\EndOfBibitem
\bibitem[Lidzey \latin{et~al.}(2000)Lidzey, Bradley, Armitage, Walker, and
  Skolnick]{Lidzey2000}
Lidzey,~D.~G.; Bradley,~D.~C.; Armitage,~A.; Walker,~S.; Skolnick,~M.~S.
  Photon-Mediated Hybridization of Frenkel Excitons in Organic Semiconductor
  Microcavities. \emph{Science} \textbf{2000}, \emph{288}, 1620--1623\relax
\mciteBstWouldAddEndPuncttrue
\mciteSetBstMidEndSepPunct{\mcitedefaultmidpunct}
{\mcitedefaultendpunct}{\mcitedefaultseppunct}\relax
\EndOfBibitem
\bibitem[Dutta \latin{et~al.}(2023)Dutta, Tiainen, Duarte, Markesevic, Morozov,
  Qureshi, Groenhof, and Toppari]{Dutta2023}
Dutta,~A.; Tiainen,~V.; Duarte,~L.; Markesevic,~N.; Morozov,~D.;
  Qureshi,~H.~A.; Groenhof,~G.; Toppari,~J.~J. Ultra-fast photochemistry in the
  strong light-matter coupling regime. \emph{Chemrxiv} \textbf{2023}, \relax
\mciteBstWouldAddEndPunctfalse
\mciteSetBstMidEndSepPunct{\mcitedefaultmidpunct}
{}{\mcitedefaultseppunct}\relax
\EndOfBibitem
\bibitem[Staroverov \latin{et~al.}(2003)Staroverov, Scuseria, Tao, and
  Perdew]{Staroverov2003}
Staroverov,~V.~N.; Scuseria,~G.~E.; Tao,~J.; Perdew,~J.~P. Comparative
  assessment of a new nonempirical density functional: Molecules and
  hydrogen-bonded complexes. \emph{J. Chem. Phys.} \textbf{2003}, \emph{119},
  12129--12137\relax
\mciteBstWouldAddEndPuncttrue
\mciteSetBstMidEndSepPunct{\mcitedefaultmidpunct}
{\mcitedefaultendpunct}{\mcitedefaultseppunct}\relax
\EndOfBibitem
\bibitem[Picconi(2021)]{Picconi2021}
Picconi,~D. Nonadiabatic quantum dynamics of the coherent excited state
  intramolecular proton transfer of 10-hydroxybenzo [h] quinoline.
  \emph{Photochem. Photobiol. Sci.} \textbf{2021}, \emph{20}, 1455--1473\relax
\mciteBstWouldAddEndPuncttrue
\mciteSetBstMidEndSepPunct{\mcitedefaultmidpunct}
{\mcitedefaultendpunct}{\mcitedefaultseppunct}\relax
\EndOfBibitem
\end{mcitethebibliography}

\end{document}

\setlength{\marginparwidth}{3cm} 
\newpage
\tableofcontents

\newpage
\section{Molecular Dynamics Simulation Model}

\subsection{Multiscale Tavis-Cummings Hamiltonian}

\begin{figure*}[!htb]
\centerline{
  \epsfig{figure=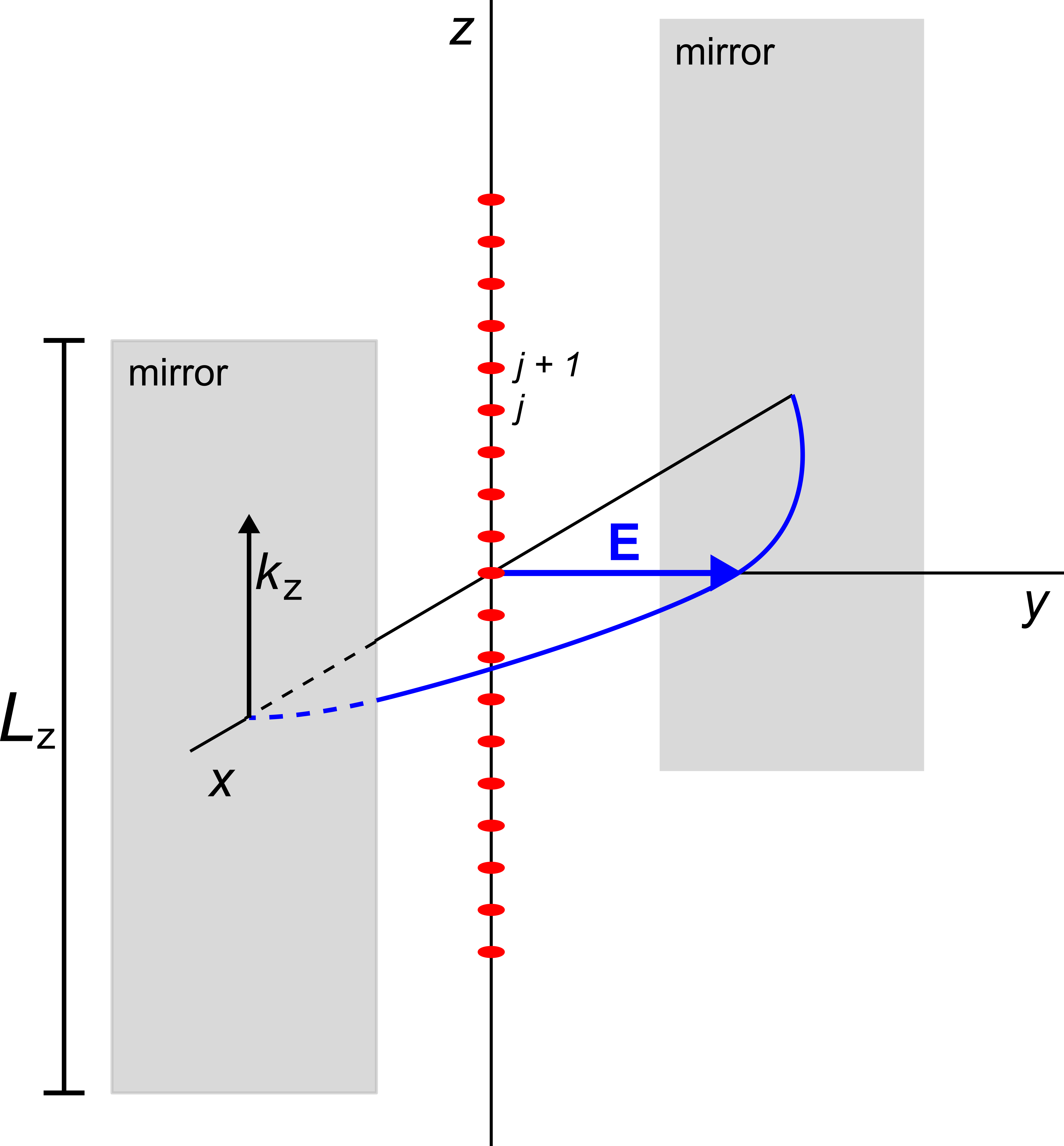,width=7cm,angle=0}}
  \caption{One-dimensional (1D) Fabry-P\'{e}rot micro-cavity model. Two reflecting mirrors located at $-\frac{1}{2}x$ and $\frac{1}{2}x$, confine light modes along this direction, while free propagation along the $z$ direction is possible for plane waves with in-plane momentum $k_z$ and energy $\hslash\omega_{\text{cav}}(k_z)$. The vacuum field vector (red) points along the $y$-axis, reaching a maximum amplitude at $x=0$ where the $N$ molecules (magenta ellipses) are placed, distributed along the $z$-axis at positions $z_j$ with $1\le j \le N$. 
  }
  \label{fig:1D_cavity}
\end{figure*}

Within the single-excitation subspace, which is probed experimentally under weak driving conditions, and employing the rotating wave approximation (RWA), which is valid for light-matter coupling strengths below 10\% of the material excitation energy,\cite{Forn-Diaz2019} the interaction between $N$ molecules and $n_\text{mode}$ confined light modes in a one-dimensional (1D)  Fabry-P\'{e}rot cavity (Figure~\ref{fig:1D_cavity}) is modelled with the multi-scale Molecular Dynamics (MD) extension of the traditional Tavis-Cummings model of quantum optics:\cite{Jaynes1963,Tavis1969,Michetti2005,Tichauer2021}
%The multi-scale Molecular Dynamics (MD) method used in this work is based on the Tavis-Cummings Hamiltonian of Quantum Optics, and models the interaction between $N$ molecules and $n_\text{mode}$ confined light modes in a one-dimensional  Fabry-P\'{e}rot cavity (Supplementary Figure~\ref{fig:1D_cavity}):\cite{Jaynes1963,Tavis1969,Michetti2005,Tichauer2021}
\begin{equation}
\begin{array}{ccl}
    {\hat{H}}^{\text{TC}} &=& \sum_j^N h\nu_j(\mathbf R_j)\hat{\sigma}^+_j\hat{\sigma}^-_j+\sum_{k_z}^{n_\text{mode}}\hslash\omega_{\text{cav}}(k_z)\hat{a}_{k_z}^\dagger\hat{a}_{k_z}+\\
    \\
    &&\sum_j^N\sum_{k_z}^{n_\text{mode}}\hslash g_j(k_z) \left(\hat{\sigma}^+_j\hat{a}_{k_z}e^{ik_zz_j}+\hat{\sigma}^-_j\hat{a}_{k_z}^\dagger e^{-ik_zz_j}\right)+\\
    \\
    &&\sum_{i}^N V^\text{mol}_{\text{S}_0}({\bf{R}}_i)
    %+\sum_i^N T({\bf{P}}_j)
    \end{array}
    \label{eq:dTCH}
\end{equation}
Here, $\hat{\sigma}_j^+$ ($\hat{\sigma}_j^-$) is the operator that excites (de-excites) molecule $j$ from the electronic ground (excited) state $|\text{S}_0^j({\bf{R}}_j)\rangle$ ($|\text{S}_1^{j}({\bf{R}}_j)\rangle$) to the electronic excited (ground) state $|\text{S}_1^{j}({\bf{R}}_j)\rangle$ ($|\text{S}_0^j({\bf{R}}_j)\rangle$); ${\bf{R}}_j$ is the vector of the Cartesian coordinates of all atoms in molecule $j$, centered at $z_j$;
$\hat{a}_{k_z}$ ($\hat{a}_{k_z}^\dagger$) is the annihilation (creation) operator of an excitation of a cavity mode with wave-vector $k_z$; $h\nu_j(\mathbf R_j)$ is the excitation energy of molecule $j$, defined as: 
\begin{equation}
  h\nu_j(\mathbf R_j)=V_{\text{S}_1}^{\text{mol}}({\bf{R}}_j)-V_{\text{S}_0}^{\text{mol}}({\bf{R}}_j)
  \label{eq:Hnu}
\end{equation}
with $V_{\text{S}_0}^{\text{mol}}({\bf{R}}_j)$ and $V_{\text{S}_1}^{\text{mol}}({\bf{R}}_j)$ the adiabatic potential energy surfaces (PESs) of molecule $j$ in the electronic ground (S$_0$) and excited (S$_1$) state, respectively. 
The last term in Equation~\ref{eq:dTCH} is the total potential energy of the system in the absolute ground state ({\it i.e.}, with no excitations in neither the molecules nor the cavity modes), defined as the sum of the ground-state potential energies of all molecules in the cavity. The $V_{\text{S}_0}^{\text{mol}}({\bf{R}}_j)$ and $V_{\text{S}_1}^{\text{mol}}({\bf{R}}_j)$ adiabatic PESs are modelled at the hybrid quantum mechanics / molecular mechanics (QM/MM) level of theory.\cite{Warshel1976b,Boggio-Pasqua2012}

% figure 1S

The third term in Equation~\ref{eq:dTCH} describes the light-matter interaction within the dipolar approximation through $g_j(k_z)$:
\begin{equation}
    g_j(k_z) = -{\boldsymbol{\upmu}}^\text{TDM}_j({\bf{R}}_j) \cdot {\bf{u}}_{\text{cav}} \sqrt{\frac{\hslash\omega_{\text{cav}}(k_z)}{2\epsilon_0 V_{\text{cav}}}} 
    \label{eq:dipole_coupling}
\end{equation}
where ${\boldsymbol{\upmu}}_j^\text{TDM}({\bf{R}}_j)$ is the transition dipole moment of molecule $j$ that depends on the molecular geometry (${\bf{R}}_j)$; ${\bf{u}}_\text{cav}$ the unit vector in the direction of the electric component of cavity vacuum field ({\it{i.e.}}, $|{\bf{E}}|=\sqrt{\hslash\omega_\text{cav}(k_z)/2\epsilon_0V_\text{cav}}$), here along the $y$-direction (Figure~\ref{fig:1D_cavity}); $\epsilon_0$ the vacuum permittivity; and $V_{\text{cav}}$ the cavity mode volume.

\subsection{One-dimensional Periodic Cavity}

Following Michetti and La Rocca,~\cite{Michetti2005} we impose periodic boundary conditions in the $z$-direction of our cavity, and thus restrict the wave vectors, $k_z$, to discrete values: $k_{z,p}=2\pi p/L_z$ with $p \in \mathbb Z$ and $L_{z}$ the length of the 1D cavity. With this approximation the molecular Tavis-Cummings Hamiltonian in Equation~\ref{eq:dTCH} can be represented as an $(N+n_{\text{mode}})$ by $(N+n_{\text{mode}})$ matrix with four blocks:\cite{Tichauer2021}
\begin{equation}
  {\bf{H}}^{\text{TC}} = \left(\begin{array}{cc} 
  {\bf{H}}^{\text{mol}} & {\bf{H}}^{\text{int}}\\
  {\bf{H}}^{\text{int}\dagger} & {\bf{H}}^{\text{cav}}\\
  \end{array}\right)\label{eq:TavisCummings}
\end{equation}
We compute the elements of this matrix in the product basis of adiabatic molecular states times cavity mode excitations:
\begin{equation}
\begin{array}{ccl}
|\phi_j\rangle &=& \hat{\sigma}_j^+|\text{S}_0^1\text{S}_0^2..\text{S}_0^{N-1}\text{S}_0^N\rangle\otimes|00..0\rangle\\
\\
&=&\hat{\sigma}_j^+|\Pi_i^N\text{S}_0^i\rangle\otimes|\Pi_k^{n_\text{mode}}0_k\rangle \\
\\
&=&\hat{\sigma}_j^+|\phi_0\rangle
\end{array}\label{eq:basis1}
\end{equation}
for $1\le j\le N$, and 
\begin{equation}
\begin{array}{ccl}
|\phi_{j >N}\rangle &=& \hat{a}_{j-N}^\dagger|\text{S}_0^1\text{S}_0^2..\text{S}_0^{N-1}\text{S}_0^N\rangle\otimes|00..0\rangle\\
\\
&=&\hat{a}_{j-N}^\dagger|\Pi_i^N\text{S}_0^i\rangle\otimes|\Pi_k^{n_\text{mode}}0_k\rangle \\
\\
&=&\hat{a}_{j-N}^\dagger|\phi_0\rangle
\end{array}\label{eq:basis2}
\end{equation}
for $N < j\le N+n_\text{mode}$. In these expressions $|00..0\rangle$ indicates that the Fock states for all $n_\text{mode}$ cavity modes are empty. The basis state $|\phi_0\rangle$ is the ground state of the molecule-cavity system with no excitations of neither the molecules nor cavity modes:
\begin{equation}
|\phi_0\rangle = |\text{S}_0^1\text{S}_0^2..\text{S}_0^{N-1}\text{S}_0^N\rangle\otimes|00..0\rangle
=|\Pi_i^N\text{S}_0^i\rangle\otimes|\Pi_k^{n_\text{mode}}0_k\rangle\label{eq:phi0}
\end{equation}

The upper left block, ${\bf{H}}^{\text{mol}}$, is an $N\times N$ matrix containing the single-photon excitations of the molecules. Because we neglect direct excitonic interactions between molecules, this block is diagonal, with elements labeled by the molecule indices $j$:
\begin{equation}
    H_{j,j}^{\text{mol}}=\langle \phi_0|\hat{\sigma}_j\hat{H}^{\text{TC}}\hat{\sigma}_j^+|\phi_0\rangle\label{eq:molecular_diagonal}
\end{equation}
for $1 \le j \le N$. 
Each matrix element of ${\bf{H}}^\text{mol}$ thus represents the potential energy of a molecule, $j$, in the electronic excited state $|\text{S}_1^{j}({\bf{R}}_j)\rangle$ while all other molecules, $i\neq j$, are in the electronic ground state $|\text{S}_0^i({\bf{R}}_i)\rangle$:
\begin{equation}  H_{j,j}^{\text{mol}}=V_{\text{S}_1}^{\text{mol}}({\bf{R}}_j)+\sum^N_{i\neq j}V_{\text{S}_0}^{\text{mol}}({\bf{R}}_i)
  \label{eq:Hj}
\end{equation}

The lower right block,  ${\bf{H}}^{\text{cav}}$, is an $n_\text{mode}\times n_\text{mode}$ matrix (with $n_{\text{mode}}=n_{\text{max}}-n_{\text{min}}+1$) containing the single-photon excitations of the cavity modes, and is also diagonal:
\begin{equation}
    H_{p,p}^{\text{cav}}= \langle \phi_0|\hat{a}_p\hat{H}^{\text{TC}}\hat{a}^\dagger_p|\phi_0\rangle\label{eq:photonic_diagonal}
\end{equation}
for $n_{\text{min}}\le p \le n_{\text{max}}$. Here, $\hat{a}_p^\dagger$ excites cavity mode $p$ with wave-vector $k_{z,p}= 2\pi p/L_z$. In these matrix elements, all molecules are in the electronic ground state (S$_0$). The energy is therefore the sum of the cavity energy at $k_{z,p}$, and the molecular ground state energies:
\begin{equation}
    H_{p,p}^{\text{cav}}= \hslash\omega_{\text{cav}}(2\pi p/L_z)+\sum^N_{j}V_{\text{S}_0}^{\text{mol}}({\bf{R}}_j)
    \label{eq:Hn}
\end{equation}
where, $\omega_{\text{cav}}(k_{z,p})$ is the cavity dispersion (dashed curve in Figure 1f, main text): 
\begin{equation}
  \omega_\text{cav}(k_{z,p})=\sqrt{\omega^2_0+ c^2k_{z,p}^2/n^2}
  \label{eq:dispersion}
\end{equation}
with $\hslash\omega_0$ the energy at $k_0=0$, $n$ the refractive index of the medium and $c$ the speed of light in vacuum.

The two $N\times n_\text{mode}$ off-diagonal blocks ${\bf{H}}^{\text{int}}$ and ${\bf{H}}^{\text{int}\dagger}$ in the multi-mode Tavis-Cummings Hamiltonian (Equation~\ref{eq:TavisCummings}) model the light-matter interactions between the molecules and the cavity modes. Within the long-wavelength approximation these matrix elements can be approximated by the overlap between the transition dipole moment of molecule $j$ and the transverse electric field of cavity mode $p$ at the geometric center $z_j$ of that molecule:
\begin{equation}
\begin{array}{ccl}
  H_{j,p}^{\text{int}} &=& 
  -{\boldsymbol{\upmu}}_j^\text{TDM}({\bf{R}}_j)\cdot {\bf{u}}_{\text{cav}} \sqrt{\frac{\hslash\omega_{\text{cav}}(2\pi p/L_z)}{2\epsilon_0 V_{\text{cav}}}} \langle\phi_0|\hat{\sigma}_j\hat{\sigma}^+_j\hat{a}_p e^{i 2\pi p z_j/L_z}\hat{a}_p^\dagger|\phi_0\rangle \\
  \\
  &=&
  -{\boldsymbol{\upmu}}_j^\text{TDM}({\bf{R}}_j) \cdot {\bf{u}}_{\text{cav}} \sqrt{\frac{\hslash\omega_{\text{cav}}(2\pi p/L_z)}{2\epsilon_0 V_{\text{cav}}}} e^{i 2\pi p z_j/L_z}     
  \end{array}
  \label{eq:QMMMdipole_coupling}
\end{equation}
for $1 \le j \le N$ and $n_{\text{min}}\le p \le n_{\text{max}}$. 

\subsection{Ehrenfest dynamics}

% propagator of the wave function

In our simulations classical trajectories evolve under the influence of the expectation value of forces with respect to the polaritonic wave function,\cite{Ehrenfest1927} while the polaritonic wave function evolves along with the classical degrees of freedom. By expanding the total wave function as a linear combination of the \emph{time-independent} diabatic light-matter states (Equations~\ref{eq:basis1} and~\ref{eq:basis2}):
\begin{equation}
|\Psi(t)\rangle = \sum_j^{N+n_\text{modes}}|\phi_j\rangle c_j(t)\label{eq:totalwf}
\end{equation}
the evolution of the \emph{time-dependent} diabatic expansion coefficients, $c_j(t)$, is obtained by numerically integrating the Schr\"{o}dinger equation over discrete time intervals, $\Delta t$. 
\begin{equation}
{\bf{c}}(t+\Delta t)={\bf{P}}^\text{dia}{\bf{c}}(t)\label{eq:SE_dia}
\end{equation}
Here, ${\bf{c}}(t)$ is a vector containing the diabatic expansion coefficients $c_j(t)$ and ${\bf{P}}^\text{dia}$ the propagator in the diabatic basis
\begin{equation}
    {\bf{P}}^\text{dia} =  \exp\left[-i\left({\bf{H}}^\text{TC}(t+\Delta t)+{\bf{H}}^\text{TC}(t)-i\hbar {\boldsymbol{\gamma}}\right)\Delta t/2\hbar\right]\label{eq:propdia}
\end{equation}
To account for the losses due to photon-leakage through the imperfect cavity mirrors, we add the decay rates of the cavity modes to the Tavis-Cummings Hamiltonian, here represented as a diagonal matrix, ${\boldsymbol{\gamma}}$, with the cavity mode decay rates $\gamma_k$ as elements.\cite{Felicetti2020,Ulusoy2020,Antoniou2020, Hu2022} Because of these decay terms, the norm of the total wave function, $|\Psi(t)|^2=\sum_j^{N+n_\text{modes}}|c_j|^2$, is not conserved but decreases due to the losses.

\section{Simulation details} 
\label{subsec:sim_details}

\subsection{HBQ model}

The Gromos-2016H66 force field was used to model the interactions, because it contains a validated cyclohexane model,\cite{Horta2016} which was used as the solvent for HBQ in our simulations. In this united-atom representation of cyclohexane, none of the atoms carries a partial charge. Because HBQ was kept frozen during the solvent equilibration and modeled at the QM level in all other simulations, there was no need for assigning partial charges to the HBQ atoms. The Gromos96 atom-types used for HBQ to model the Van der Waals interactions with the cyclohexane solvent, are {\tt{HC}} for all aromatic hydrogen atoms, {\tt{C}} for all carbon atoms, {\tt{NR}} for the aromatic nitrogen atom, {\tt{OA}} for the hydroxyl oxygen atom and {\tt{H}} for the hydroxyl proton.

One HBQ molecule was geometry-optimized and placed inside a rectangular box that was subsequently filled with 402 cyclohexane molecules. The simulation box, which thus contained 2436 atoms, was equilibrated for 100 ns. During equilibration, the coordinates of the HBQ atoms were kept fixed. The LINCS algorithm was used to constrain bond lengths in cyclohexane,\cite{Hess1997} enabling a time step of 2~fs. Temperature and pressure were maintained at 300~K and 1~atmosphere by means of weak-coupling to an external bath ($\tau_T =$~0.1 ps, $\tau_P = $~1 ps).\cite{Berendsen1984} The Lennard-Jones potential was truncated at 1.4 nm. 

Snapshots were extracted from the equilibration trajectory and further equilibrated for 50 ps at the QM/MM level with a time step of 1 fs. In these QM/MM simulations HBQ was modelled with density functional theory (DFT), using the CAM-B3LYP functional,\cite{Becke1993,Lee1988,Yanai2004} in combination with a 3-21G basis set.\cite{Dunning1970} The cyclohexane solvent was described with the 2016H66 parameter set of the Gromos96 force field.\cite{Horta2016} The QM subsystem was mechanically embedded within the MM subsystem. Because cyclohexane atoms are uncharged, the interactions between the QM and MM regions were modeled with Lennard-Jones potentials only. We used time-dependent DFT (TD-DFT),\cite{Runge1984} within the Tamm-Dancoff approximation (TDA),\cite{Hirata1999} in combination with the CAM-B3LYP functional and the 3-21G basis set, to model the singlet excited electronic (S$_1$) state of HBQ. All QM/MM simulations of HBQ outside of the cavity were performed with GROMACS version~4.5.3,\cite{Hess2008} using the QM/MM interface to  TeraChem version~1.93.\cite{Ufimtsev2009,Titov2013}

\subsection{Methylene Blue model}

\begin{figure*}[!tb]
\centering
\includegraphics[width=1\textwidth]{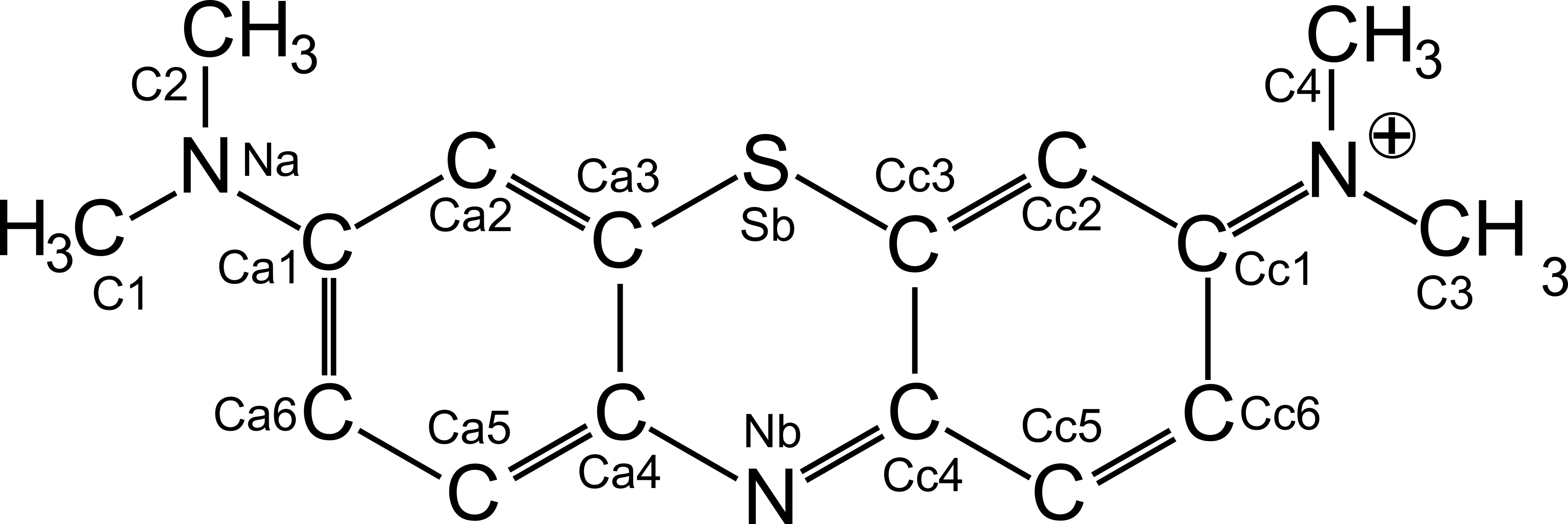}
  \caption{Structure and atom names of Methylene Blue. Not all hydrogens are shown, but their names take the name of the carbon atom plus an integer, {\it e.g.} HA2 is attached to CA2 and H21, H22 and H23 are attached to carbon C2.}
\label{fig:MeB}
\end{figure*}

The Amber03 force field was used to model the interactions between Methylene Blue (MeB) and the water solvent, which was described with the TIP3P water model.\cite{Jorgensen1983} Atom types and partial charges for the MeB atoms (Figure~\ref{fig:MeB}) are listed in Table~\ref{tab:MeB}. The partial charges of the atoms were derived following the procedure recommended for the Amber03~force field.\cite{Duan2003,Bayly1993} First, the geometry of MeB was minimized at the HF/6-31G** level of \textit{ab initio} theory, using the IEFPCM continuum solvent model with a relative dielectric constant of 4.0.\cite{Tomasi2005} After the geometry optimizations, the electrostatic potential at 10~concentric layers of 17~points per unit area around each atom was evaluated using the electron density calculated at the B3LYP/cc-pVTZ level of DFT theory,\cite{Becke1993} again using the IEFPCM continuum solvent model with a relative dielectric constant of 4.0.\cite{Tomasi2005} The atomic charges were obtained by performing a two-stage RESP fit to the electrostatic potential,\cite{Bayly1993} the first without symmetry constraints, and the second with symmetry constraints on chemically identical atoms.

\begin{table}[!htb]
\caption{Amber03 atomtypes and partial charges for Methylene Blue. Atom names are defined in Figure~\ref{fig:MeB}.}
\label{tab:MeB}
\begin{tabular}{l|l|r}
\hline
name  & type & charge (e)\\
\hline
  Ca1 & CA  &    0.054312 \\
   Na & NA  &    0.154719 \\
   C1 & CT  &   -0.284923 \\
  H11 & H1  &    0.128201 \\
  H12 & H1  &    0.128201 \\
  H13 & H1  &    0.128201 \\
   C2 & CT  &   -0.275632 \\
  H21 & H1  &    0.127095 \\
  H22 & H1  &    0.127095 \\
  H23 & H1  &    0.127095 \\
  Ca2 & CA  &   -0.205067 \\
  Ha2 & HA  &    0.181067 \\
  Ca3 & C   &   -0.101342 \\
  Ca4 & CA  &    0.624779 \\
  Ca5 & CA  &   -0.286158 \\
  Ha5 & HA  &    0.187369 \\
  Ca6 & CA  &   -0.141958 \\
  Ha6 & HA  &    0.158598 \\
   Nb & NB  &   -0.712421 \\
   Sb & S   &    0.049117 \\
  Cc1 & CA  &    0.054312 \\
   Nc & NA  &    0.154719 \\
   C3 & CT  &   -0.284923 \\
  H31 & H1  &    0.128201 \\
  H32 & H1  &    0.128201 \\
  H33 & H1  &    0.128201 \\
   C4 & CT  &   -0.275632 \\
  H41 & H1  &    0.127095 \\
  H42 & H1  &    0.127095 \\
  H43 & H1  &    0.127095 \\
  Cc2 & CA  &   -0.205067 \\
  Hc2 & HA  &    0.181067 \\
  Cc3 & C   &   -0.101342 \\
  Cc4 & CA  &    0.624779 \\
  Cc5 & CA  &   -0.286158 \\
  Hc5 & HA  &    0.187369 \\
  Cc6 & CA  &   -0.141958 \\
  Hc6 & HA  &    0.158598 \\
\hline
\end{tabular}
\end{table}

A single MeB molecule was geometry-optimized at the B97/3-21G level of theory and placed at the center of a rectangular box that was filled with 2031 TIP3P water molecules.\cite{Jorgensen1983} A 1.0~nm cut-off was used for the Van der Waals' interactions, which were modeled with Lennard-Jones potentials, while the Coulomb interactions were computed with the smooth particle mesh Ewald method,\cite{Essmann1995} using a 1.0~nm real space cut-off and a grid spacing of 0.12~nm and a relative tolerance at the real space cut-off of 10$^{-5}$. The simulation box, containing 6,131 atoms, was equilibrated for 1~ns at the force field level of theory. During this equilibration, the coordinates of the MeB atoms were kept fixed. The temperature was kept at 300~K with the v-rescale thermostat,\cite{Bussi2007} while the pressure was kept constant at 1~atmosphere using the Berendsen isotropic pressure coupling algorithm,\cite{Berendsen1984} with a time constant of 1~ps. The SETTLE algorithm was applied to constrain the internal degrees of freedom of water molecules,\cite{Miyamoto1992} enabling a time step of 2~fs in the classical MD simulations. 

After equilibration at the force field (MM) level, the system was further equilibrated at the QM/MM level for 10~ps. The time step was reduced to 1~fs. The Methylene Blue molecule was modelled at the DFT level, using the B97 functional,\cite{Becke97} in combination with the 3-21G basis set.\cite{Dunning1970} The water solvent was modelled with the TIP3P force field.\cite{Jorgensen1983} The QM system experienced the Coulomb field of all MM atoms within a 1.0~nm cut-off sphere and Lennard-Jones interactions between MM and QM atoms were added. The singlet electronic excited state (S$_1$) was modeled with TD-DFT,\cite{Runge1984} using the B97 functional in combination with the 3-21G basis set for the QM region ({\it {i.e.,}} TD-B97/3-21G),\cite{Dunning1970}. At this level of QM/MM theory, the excitation energy is 2.5~eV. The overestimation of the vertical excitation energy is due to the limited accuracy of the employed level of theory, but was compensated by adding an off-set to the cavity resonance energy (see subsection~\ref{subsec:cavitymd}). The QM/MM simulations were performed with GROMACS version 4.5.3,\cite{Hess2008} interfaced to TeraChem version 1.93.\cite{Ufimtsev2009,Titov2013}

\subsection{HBQ/MeB cavity simulations}\label{subsec:cavitymd}

From the HBQ QM/MM trajectory, we selected single HBQ snapshots, which we combined with 1023~Methylene Blue molecules from the MeB QM/MM trajectory. These 1024~molecules, including their solvent environments, were placed at equal inter-molecular separations on the $z$-axis of a 1D,\cite{Michetti2005} 50~$\upmu$m long, symmetric optical Fabry-P\'erot micro-cavity. The dispersion of this cavity was modelled  with 239~discrete modes ({\it i.e.}, $k_{z,p}=2\pi p/L_z$ with $-119\leq p\leq 119$ and $L_z=$~50 $\upmu$m). The micro-cavity was red-detuned with respect to the Methylene Blue absorption maximum, which is 2.5~eV at the TD-B97/3-21G//Amber03 level of theory used in this work, such that the energy of the fundamental cavity mode at zero incidence angle ($k_0=0$) was $\hslash\omega_0 =$~2.18~eV. To maximize the collective light-matter coupling strength, the transition dipole moments of all molecules were aligned to the vacuum field at the start of the simulation. With a vacuum field strength of ${\bf{E}}_y =$ 0.00004 (0.21~MVcm$^{-1}$) this configuration yields a Rabi splitting of $\hbar\Omega^\text{Rabi}= $~323~meV. All simulations were performed for a lossy cavity with a decay rate of $\hslash\gamma_{\text{cav}}=$ 0.04~eV or $\gamma_{\text{cav}}=$ 66.7~ps$^{-1}$. At such rate, the lifetime, $\tau_\text{cav}$, of the lossy cavity is comparable to the 2-14~fs lifetimes of metallic Fabry-P\'{e}rot cavities used in experiments on strong coupling with organic molecules.\cite{Schwartz2013,George2015,Rozenman2018} With a lifetime of 15~fs and a resonance at 2.18~eV, the Quality-factor, defined as $\text{Q}=\omega_\text{cav}\tau_\text{cav}$, would be 50 for our cavity. A leap-frog Verlet algorithm with a 0.5~fs time step was used to integrate the nuclear dynamics, while the unitary propagator in the diabatic basis was used to propagate the polaritonic degrees of freedom.\cite{Granucci2001} The simulation was performed with GROMACS version~4.5.3,\cite{Hess2008} in which the multi-mode Tavis-Cummings QM/MM model was implemented,\cite{Tichauer2021} in combination with Gaussian16.\cite{g16} The GROMACS source code is available for download from {\tt{https://github.com/upper-polariton/GMXTC}}. The results presented in the main manuscript are averages over four trajectories, started  with different initial conditions.

% Trying to avoid silly questions about using different solvents and MM descriptions.

Because the molecules do not interact directly, but rather via the cavity modes, there are no issues in using different QM/MM descriptions for HQB and MeB. Although the solvents as well as the force fields and QM methods were chosen because of convenience, we emphasize that for the purpose of this work it is not essential to have the most accurate description of the bare excitation energies of the molecules, but rather to have a realistic model of the molecular degrees of freedom, including the solvent environment. We speculate that for practical realizations HBQ could be embedded via small micro-droplets of a suitable solvent within the polymer matrix containing MeB, or \textit{vice versa}.

\section{Simulation Analysis}

\subsection{Monitoring exciton dynamics in the strong coupling regime}
\label{sec:analysis-wps}

To track the propagation of the excitonic wavepacket, we plotted the probability density of the excitonic contributions to the total time-dependent wave function $|\Psi(t)|^2$ at the positions of the molecules,  $z_j$, as a function of time. We thus represent the density as a \emph{discrete} distribution at grid points that correspond to the molecular positions, rather than a continuous distribution. 
%To obtain the total probability density of the excitonic wave function, $|\Psi-\text{exc}(t)|^2$, we first computed the probability densities of the excitonic $|\Psi_{\text{exc}}(t)|^2$ and photonic $|\Psi_{\text{pho}}(t)|^2$ contributions, and then summed these contributions. 
The amplitude of the excitonic wave function, $|\Psi_{\text{exc}}(t)\rangle$, at position $z_j$ in the 1D cavity (with $z_j=(j-1)L_z/N$ for $1\le j \le N$) is obtained by projecting the excitonic basis state in which molecule $j$ at position $z_j$ is excited ($|\phi_j\rangle=\hat{\sigma}_j^+|\phi_0\rangle$) to the total wave function (Equation~\ref{eq:totalwf}):
\begin{equation}
    |\Psi_{\text{exc}}(z_j,t)\rangle=\hat{\sigma}_j^+|\phi_0\rangle
        \langle\phi_0|\hat{\sigma}_j|\Psi(t)\rangle = c_j(t)\label{eq:molecularpart}
\end{equation}
where the $c_j(t)$ is the time-dependent expansion coefficients of the total wave function $|\Psi(t)\rangle$ (Equation~\ref{eq:totalwf}).

%The cavity mode excitations are described as plane waves that are delocalized in real space. We therefore obtain the amplitude of the cavity modes at position $z_j$ by Fourier transforming the projection of the cavity mode excitation basis states, in which cavity mode $p$ is excited ($|\phi_p\rangle=\hat{a}_{p}^\dagger|\phi_0\rangle$):
%\begin{equation}
%\begin{array}{ccl}
%    |\Psi_\text{pho}(z_j,t)\rangle&=&\mathcal{FT}^{-1}\left[\sum_p^{n_\text{mode}}\hat{a}_p^\dagger|\phi_0\rangle\langle\phi_0|\hat{a}_p|
%    \Psi\rangle\right]\\
%    \\
%    &=&\frac{1}{\sqrt N}\sum_p^{n_{\text{mode}}} d_{N+p}(t)e^{i2\pi z_j p}
%    \end{array}\label{eq:photonicpart1}
%\end{equation}
%where the $d_{N+p}(t)$ are the time-dependent expansion coefficients of the cavity mode excitations of the total wave function, $|\Psi(t)\rangle$ (Equation~\ref{eq:totalwf}). Here, we normalize by $1/\sqrt{N}$ rather than $1/\sqrt{L_z}$, as we represent the density on the grid of molecular positions. 

\subsection{Photo-Absorption Spectra}
\label{sec:absorption}

\subsubsection{Molecular absorption spectra}

The absorption spectra of HBQ and MeB were computed from QM/MM trajectories in the ground (S$_0$) state. These spectra were composed of the S$_1$-S$_0$ energy gaps by super-position of Gaussian functions:\cite{Modi2019}
\begin{equation}
  I^{\text{abs}}_{\text{mol}}(E) \propto \sum_i^s \Delta E_i\exp\left[-\frac{(E-\Delta
    E_i)^2}{2\sigma^2}\right]|{\boldsymbol{\mu}}^{\text{TDM}}_i|^2\label{eq:abs_mol}
\end{equation}
where $I^{\text{abs}}_{\text{mol}}(E)$ is the intensity of the molecular absorption as a function of excitation energy ($E$), $s$ the number of snapshots included in the analysis, $\Delta E_i$ the excitation energy in snapshot $i$ and ${\boldsymbol{\mu}}_i^\text{TDM}$ the transition dipole moment of that excitation. A width of $\sigma = 0.05$~eV was used for the convolution.

The emission spectrum of HBQ was computed from a 10~ps QM/MM trajectory in the excited (S$_1$) state. The first 100~fs were discarded to sample only the relaxed photo-product state. The spectra were composed as in Equation~\ref{eq:abs_mol}, using the S$_1$-S$_0$ energy gaps and transition dipole moments from the trajectory. As for the absorption spectra, a width of $\sigma = 0.05$~eV was used for the convolution. 

\subsubsection{Cavity absorption spectra}

To compute the absorption spectrum of the molecule-cavity system as a function of energy and $k_z$-vector, we first diagonalized the Tavis-Cummings Hamiltonian, using molecular energies and transition dipole moments from the equilibrium QM/MM simulations, to obtain the $N+n_\text{mode}$ hybrid light-matter states $\lvert\psi^m\rangle$:\cite{Michetti2005,Agranovich2007}
\begin{equation}
\vert\psi^{m}\rangle=\left(\sum_{j}^N\beta^m_j\hat{\sigma}^+_j + \sum_{p}^{n_{\text{mode}}}\alpha^m_p\hat{a}_p^\dagger\right)\vert\phi_0\rangle
%\vert\text{S}_0^1(\mathbf{R}_1)\text{S}_0^2(\mathbf{R}_2)..\text{S}_0^{N-1}(\mathbf{R}_{N-1})\text{S}_0^{N}(\mathbf{R}_N)\rangle\otimes\vert0\rangle
\label{eq:Npolariton}
\end{equation}
with eigenenergies $E_m$. The $\beta_j^m$ and $\alpha_p^m$ expansion coefficients reflect the contribution of the molecular excitations ($\vert\text{S}_1^j(\mathbf{R}_j)\rangle$) and the cavity mode excitations ($\vert 1_p\rangle$) to polariton $\vert \psi^m\rangle$. 

Following Lidzey and coworkers,\cite{Lidzey2000} we define the "visibility", $I^m$, of  polaritonic state $|\psi^m\rangle$ as the total photonic contribution to that state ({\it{i.e.}}, $I^m\propto \sum_p^{n_\text{max}}|\alpha_p^m|^2$). Thus, for each frame of the MeB QM/MM trajectory, the polaritonic states were computed and the energy gaps of these states with respect to the overall ground state ({\it{i.e.}}, $E^0$, with all molecules in S$_0$, no photon in the cavity, Equation~\ref{eq:phi0}) were extracted for all cavity modes, $p$, and summed up into a superposition of Gaussian functions, weighted by $|\alpha_p^m|^2$:
\begin{equation}
  I^{\text{abs}}(E,k_{z,p}) \propto \sum_t^s\left[ \sum_m^{N+n_\text{max}+1}|\alpha_{p,t}^m|^2\exp[-\frac{(E-\Delta
    E_t^m)^2}{2\sigma^2}]\right]\label{eq:abs_cav}
\end{equation}
Here, $I^{\text{abs}}(E,k_{z,p})$ is the absorption intensity as a function of excitation energy $E$ and in-plane momentum $k_{z,p}= 2\pi p/L_z$, $s$ the number of trajectory frames included in the analysis, $\Delta E_t^m$ the excitation energy of polaritonic state $m$ in frame $t$ ({\it{i.e.}}, $\Delta E^m_t = E^m_t-E^0_t$) and $\alpha_{p,t}^m$ the expansion coefficient of cavity mode $p$ in polaritonic state $m$ in that frame (equation~\ref{eq:Npolariton}). A width of $\sigma = 0.05$~eV was used for the convolution.

\section{Additional Results} 
\label{subsec:add_sim}

\subsection{Transparency of cavity mirrors}

% spectra of Methylene Blue from https://omlc.org/spectra/mb/mb-water.html

\begin{figure*}[!thb]
\centering
\includegraphics[width=\textwidth]
{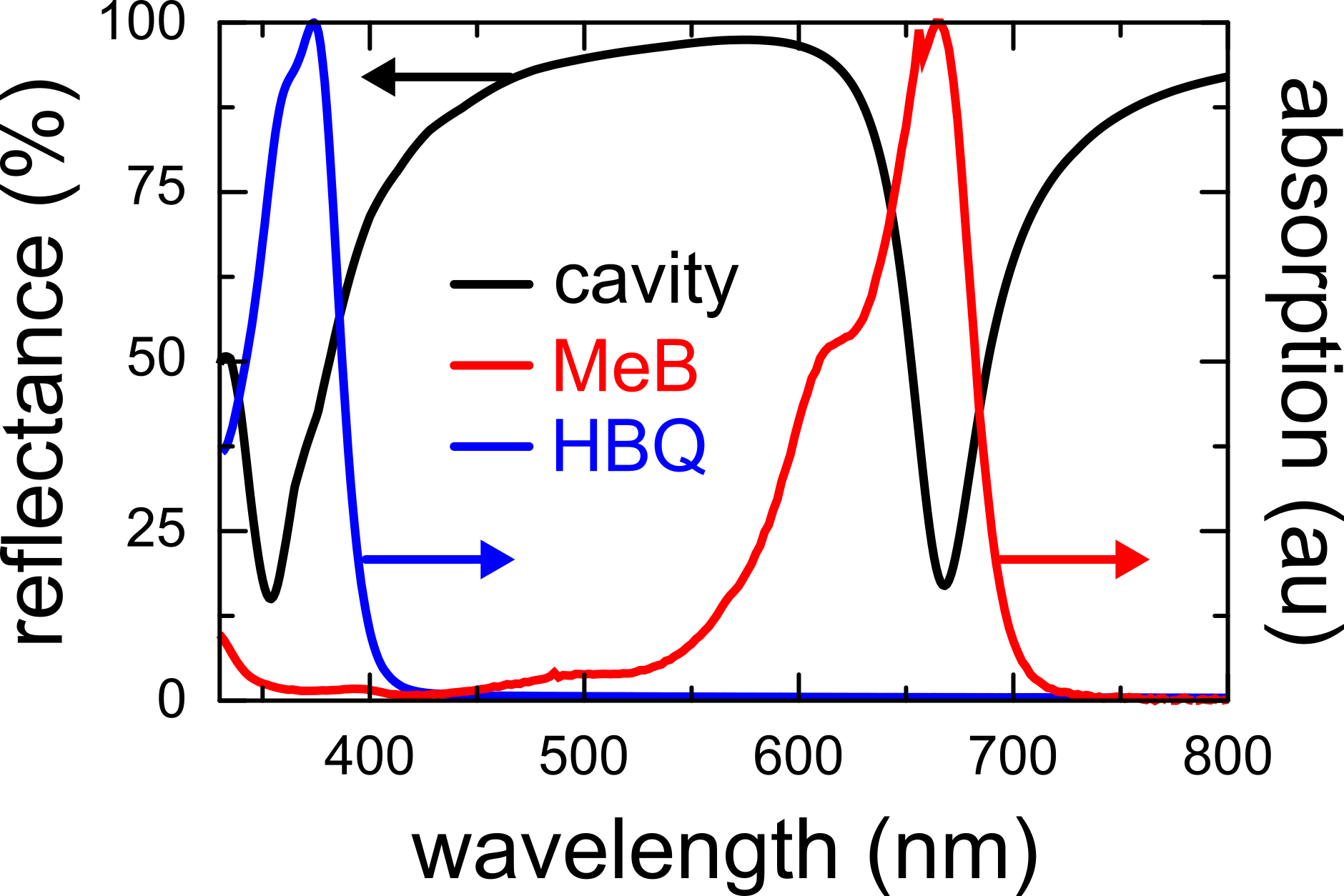}
  \caption{Calculated reflectance (black) spectrum of a Fabry-P\'{e}rot micro-cavity with a 20~nm silver mirror on top of a 167~nm polymer layer deposited on a silver substrate. Absorption spectra of MeB and HBQ are plotted in red and blue, respectively.
  }\label{fig:reflectance}
\end{figure*}

The reflectivity of a hypothetical Farby-P\'{e}rot micro-cavity composed of a 20~nm silver mirror on top of a 167~nm acrylic polymer layer and deposited on a silver substrate, was computed with the "Reflectance Calculator" of Filmetrics$^\text{\textregistered}$ ({\tt{https://www.filmetrics.com/reflectance-calculator}}) and plotted as a function of wavelength (black) in Figure~\ref{fig:reflectance}, together with the absorption spectra of HBQ in polymethyl-methacrylate (PMMA) (blue),\cite{Dutta2023}  and of MeB in water (red). Because the zeroth-order cavity mode is resonant with the main absorption peak of MeB centered at 668~nm, the cavity is transparent around that wavelength. With a sufficiently high concentration of MeB in the polymer layer, this absorption peak will split into the LP and UP bands, separated by the Rabi splitting. Because the reflectivity of the top mirror is reduced by more than 50\% at the absorption maxima of HBQ (375~nm and 360~nm), we anticipate that this mixed HBQ/MeB-cavity system will still have a significant cross-section for selectively exciting HBQ, and initiate the transport via the excited-state intra-molecular proton transfer process.

\subsection{Potential energy surface of HBQ}

\begin{figure}[!hbt]
\centering
\includegraphics[width=\textwidth]{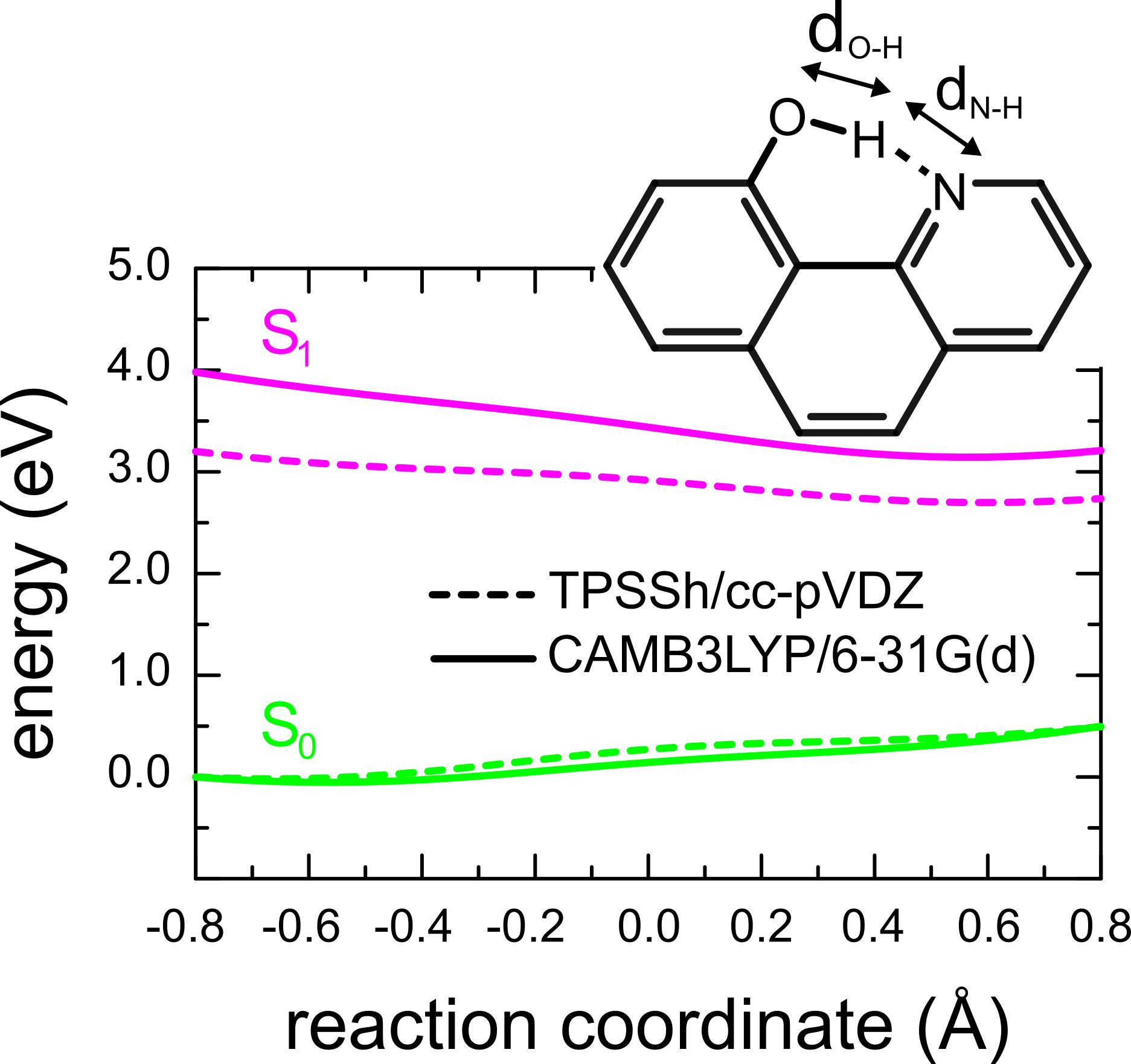}
\caption{Minimum energy profile for excited-state intra-molecular proton transfer in HBQ as a function of the reaction coordinate defined as $d_\text{O-H} - d_\text{N-H}$ (inset) in the electronic ground state (S$_0$, green) and excited state (S$_1$, magenta) at the CAMB3LYP/3-21G (continuous lines) TPSSh/cc-pVDZ (dashed lines) levels of theory.}
\label{fig:pes}
\end{figure}

In Figure~\ref{fig:pes} we compare the minimum energy potential energy profiles for the proton transfer reaction in the ground (S$_0$) and first excited electronic states (S$_1$) at the CAMB3LYP/3-21G,\cite{Yanai2004} and TPSSh/cc-pVDZ levels of theory.\cite{Staroverov2003} These profiles were obtained by performing geometry optimizations on the S$_1$ potential energy surface at 33~equidistant points along the reaction coordinate, defined as the difference between the O-H and N-H distances (inset in Figure~\ref{fig:pes}). The results of the scan suggest that the CAMB3LYP/3-21G model provides a potential energy surface on which proton transfer is barrierless in the excited state, in qualitative agreement with the calculations at the TPSSh/cc-pVDZ level of theory of Picconi,\cite{Picconi2021} that, in turn, are in good agreement with experiment.

\subsection{Delayed exciton transport when HBQ reacts later}

\begin{figure*}[!htb]
\centering
\includegraphics[width=1\textwidth]{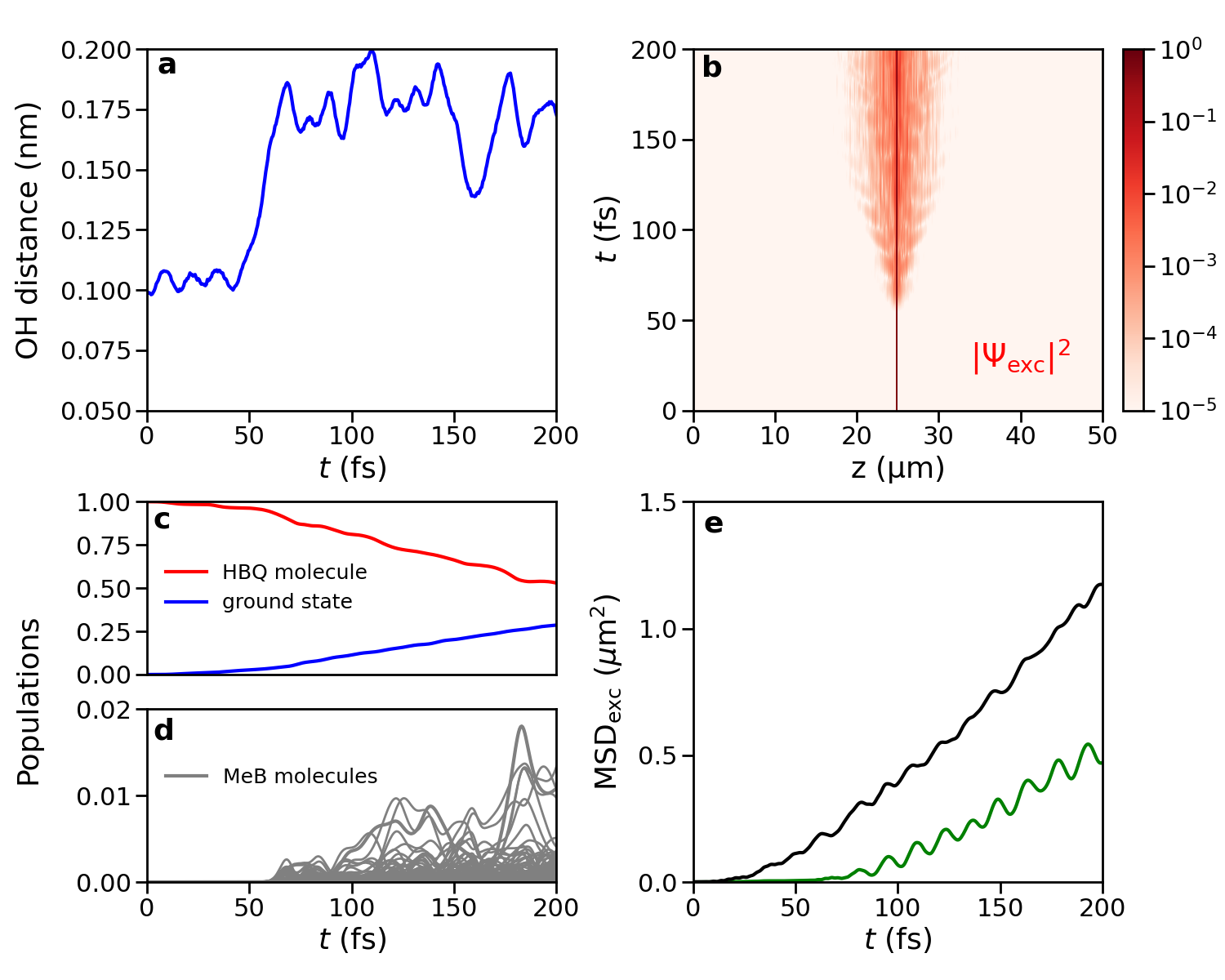}
  \caption{Delayed polariton-mediated exciton transport when the ESIPT in HBQ occurs later. Panel {\bf{a}} shows the distance between the hydroxyl oxygen and proton, which we use as the reaction coordinate for the excited state proton transfer reaction as (Figure~\ref{fig:pes}), a function of time after instantaneous excitation into the highest-energy eigenstate of the molecule-cavity system, which is dominated by the S$_1$ state of HBQ (\textit{i.e.}, $|\beta_{\text{HBQ}}|^2 \approx$ 0.99). Panel {\bf{b}} shows the probability density of the excitonic part, $|\Psi_{\text{exc}}|^2$, of the total wavefunction as a function of the $z$-coordinate (horizontal axis) and time (vertical axis). Panels {\bf{c}} and {\bf{d}} show the contribution of the HBQ molecule (red) and Methylene Blue molecules (grey) to the total wavefunction, as well as the population of the ground state (blue). Panel {\bf{e}} shows the Mean Squared Displacement (\textit{i.e.}, $\text{MSD}_{\text{exc}} = \langle z(t)-z(0)\rangle^2$) of the excitonic wavepacket depicted in Figure~3\textbf{e} of the main text (black line) and panel~\textbf{b} (green line).}
\label{fig:transport_slow}
\end{figure*}

In Figure~\ref{fig:transport_slow} we show that when ESIPT occurs later, here 50~fs after photo-excitation, also the polariton-mediated exciton transport starts later.

\section{Animations}

Animations of the excitonic wavepackets shown in Figure~3 of the main text (transport.mp4) and in Figure~\ref{fig:transport_slow} (transport\_delayed.mp4) are available for download in MPEG-4 format.

\bibliography{bibliography}